\documentclass{iopjournal}

\usepackage{amsmath, mathtools,amssymb,bm,float,mathrsfs}
\newcommand{\al}[1]{\begin{align} #1 \end{align}}
  \def\dd{\mathrm{d}}
  \def\pd{\partial}

\usepackage[caption=false]{subfig}

\newcommand{\bp}{\beta_{\rm phys}}

\renewcommand{\Re}{\operatorname{Re}}

\begin{document}

\articletype{Paper} 

\title{Stability and quasi-normal ringing in analogue black-white holes
in SNAIL-based traveling-wave parametric amplifiers}

\author{Daisuke Yamauchi$^{1,*}$, Haruna Katayama$^{2,3}$ and Norihiro Tanahashi$^4$}

\affil{$^1$Department of Physics, Faculty of Science, Okayama University of Science, 1-1 Ridaicho, Okayama, 700-0005, Japan}

\affil{$^2$Graduate School of Advanced Science and Engineering, Hiroshima University, Higashihiroshima, Hiroshima 739-8521, Japan}

\affil{$^3$Department of Physics and Astronomy, Dartmouth College, Hanover, New Hampshire 03755, USA}

\affil{$^4$Department of Physics, Kyoto University, Kyoto 606-8502, Japan}

\email{d-yamauchi@ous.ac.jp, halna496@hiroshima-u.ac.jp, tanahashi@gauge.scphys.kyoto-u.ac.jp}


\begin{abstract}
The circuit dynamics constructed by
traveling-wave parametric amplifiers (TWPA),
using superconducting nonlinear asymmetric inductive elements (SNAILs),
are known to be approximately 
described by the Korteweg-de Vries (KdV) or modified
KdV equations in the continuum limit and admit soliton solutions.
The soliton spatially modulates the effective propagation velocity of the weak probe field, 
which leads to the effective realization of the causal structure
of the analogue event horizons in the SNAIL-TWPA circuit system.
In this paper, we derive the master equation for the weak probe field
where the background soliton acts as an effective potential.
We show, in the effective theory where the nonlinear dispersion is neglected, the absence of normalizable negative modes in the SNAIL-TWPA circuit system 
by using the language of supersymmetric quantum mechanics.
We also present the first study of quasi-normal modes (QNM) 
of the SNAIL-TWPA analogue black-white hole system in the dispersionless effective theory
by numerical and semi-analytic methods.
Based on the resultant QNM frequency, we clarify the timescale 
at which nonlinear dispersion becomes effective in the SNAIL-TWPA 
circuit system and 
suggest the possibility of observing ringdown.
\end{abstract}

\section{Introduction}

Analogue black holes have been proposed in various laboratory systems such as Bose-Einstein condensates~\cite{Steinhauer:2014dra,Steinhauer:2015saa}, 
optical fibers~\cite{Philbin:2007ji,Choudhary:2012,Drori:2018ivu}, 
superfluids
in microwave cavities \cite{Nguyen:2015ilv,Jacquet:2020znq}, and electrical circuits \cite{Schutzhold:2005,Nation:2009,Katayama:2020psv,Katayama:2021ycw,Katayama:2021prd,Katayama:2021ieee,Katayama:2021itj,Katayama:2022qmr}
(see, e.g., \cite{Barcelo:2005fc} for a review).
It has been pointed out that these analogue systems 
can be used to observe quantum effects, such 
as quantum-correlated Hawking radiation.
Recently, it has been shown in 
\cite{Katayama:2022qmr}
that traveling-wave parametric amplifier (TWPA) setups
using superconducting nonlinear asymmetric inductive elements (SNAILs) 
admit soliton solutions that act as analogue event horizons.
When considering a weak probe field living on top of 
the background soliton,
the effective propagation velocity of the weak probe field can be shown to 
be modulated due to the shape of the background soliton,
which leads to the effective realization of the causal structure
of the analogue black and white holes.
It has also been suggested that this system can exhibit
interesting behaviors, such as black hole lasers \cite{Steinhauer:2015saa,Katayama:2021itj,Katayama:2022qmr,Corley:1996,Corley:1998,Corley:1999,Gaona-Reyes:2017,Faccio:2012,Leonhardt:2008}, through
nonlinear interactions and nonlinear dispersion.
In order to accurately understand these phenomena, it is crucial to
understand the behavior of the linear perturbation
that forms the basis of the interaction picture in the curved background. 
Hence, the linear perturbation theory
for the SNAIL-TWPA circuit system should be developed 
to investigate the effects of nonlinear interactions.

A realistic situation can never be fully described by 
its simple basic parameters and is always in a perturbed state.
Objects that are unstable under small perturbations will inevitably be 
destroyed by them and cannot exist.
When we would like to understand the stability of the system
on which we are focusing, we have to start with the analysis of 
the small perturbations around the background.
One method for investigating the stability of the system involves
using the effective potential with respect to the perturbations around
the background.
If the effective potential is non-negative everywhere,
it can be shown that there is no normalizable negative eigenvalue (growing) mode.
Even if the effective potential is not positive definite everywhere,
in some cases 
the special trick known as the $S$-deformation method can be used,
which allows us to prove stability (see, e.g., \cite{Kodama:2003jz,Ishibashi:2003ap,Kimura:2017uor}).

If the system is stable, once the system is perturbed, 
it eventually rings down to its final state with a long period 
of damped proper oscillation, 
which is usually determined by the so-called quasi-normal modes (QNM)~\cite{Nollert:1999ji,Kokkotas:1999bd,Berti:2009kk,Konoplya:2011qq,Hatsuda:2021gtn}.
These are eigenmodes of the evolution operator, with
a discrete complex spectrum $\Omega_n$\,, where $n$ denotes
the overtone index.
The QNM with the smallest magnitude of the imaginary part of the frequency,
namely the so-called
fundamental mode, is the longest-lived, eventually dominating
the signal.
The phenomenon of ringdown allows us to address questions
regarding the late-time dynamics and the stability of the system. Moreover, the QNM spectrum provides characteristic information about the effective background experienced by the perturbations.

Several analyses of stability and QNMs have already been 
conducted in analogue systems;
e.g., optical solitons~\cite{Burgess:2023pny}, 
Laval nozzles~\cite{Okuzumi:2007hf}, draining bathtub~\cite{Berti:2004ju} and so on. The SNAIL-TWPA platform provides a complementary and highly tunable setting for such studies. In particular, the nonlinear properties of the circuit can be tuned by an externally applied magnetic flux, which changes the available soliton solutions and their profiles, while parameters such as the soliton amplitude can be selected through the initial excitation. Since the soliton background determines the effective metric, this controllability makes the SNAIL-TWPA circuit particularly suitable for systematically investigating how the QNM spectrum changes with the effective metric and for testing the robustness of quasinormal dynamics across different analogue configurations within the same platform. Moreover, its superconducting microwave implementation may eventually provide a route toward studying QNM phenomena in regimes where quantum fluctuations and correlations become relevant. Despite these attractive features, the QNMs of this platform have not yet been studied.

In this paper, as a first step toward systematic QNM studies in the SNAIL-TWPA circuit system, we investigate the linear stability and QNMs of a soliton-induced black-white-holes within the dispersionless effective theory.
We will demonstrate that the modulation of the probe field velocity 
induced by the background soliton in the SNAIL-TWPA circuit system
provides a novel way to create an effective potential
for the probe field.
In particular, suitable perturbations to the soliton
can be shown to obey a Schrödinger-type equation with
a repulsive potential.
We will show the stability using the effective potential, and then
evaluate the QNM frequency of the SNAIL-TWPA black-white holes, 
particularly its fundamental mode.
Using the resultant expression, we will 
estimate the timescale at which nonlinear dispersion becomes effective.

This paper is organized as follows.
In Sec.~\ref{sec:2}, we first briefly review
the soliton solutions in the SNAIL-TWPA circuit system 
as background.
In Sec.~\ref{sec:3}, we derive a perturbation equation 
in which the background soliton acts as a potential,
neglecting the higher-derivative terms.
In Sec.~\ref{sec:4}, we show the absence of normalizable
negative modes corresponding to
unstable solutions in our analogue system by using the method of supersymmetric quantum mechanics.
In Sec.~\ref{sec:5}, we calculate
the complex QNM frequency, 
particularly the fundamental mode,
and identify its parameter dependence.
Section \ref{sec:conclusion} is devoted to the summary and discussion.

\section{Soliton as background}
\label{sec:2}

\begin{figure}[tbp]
    \centering
    \includegraphics[height=55mm]{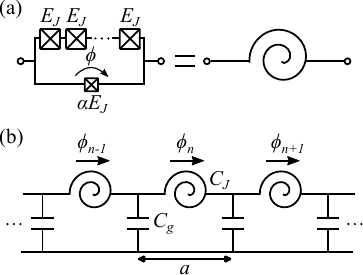}
     \caption{(a) Schematic representation of a superconducting nonlinear asymmetric inductive element (SNAIL). It consists of a superconducting loop with asymmetric Josephson junctions: one smaller junction with Josephson energy $\alpha E_J (\alpha < 1)$ and multiple larger junctions with energy $E_J$. An external magnetic flux threading the loop tunes the nonlinearity of the effective inductance. In circuit schematics, the SNAIL is represented by a spiral symbol. (b) Schematic of a SNAIL-based traveling-wave parametric amplifier (TWPA), consisting of identical SNAIL unit cells connected in series and shunted by capacitances $C_g$. Here, $C_J$ represents the effective capacitance of a SNAIL, and $a$ represents the length of a unit cell. The superconducting phase difference across the $n$-th SNAIL is denoted by $\phi_n$.}
     \label{fig:SNAIL}
\end{figure}

We first briefly review the soliton solutions serving as the background, 
following Ref.~\cite{Katayama:2022qmr}.
Based on the circuit model shown in Fig.~\ref{fig:SNAIL}, we write down the circuit equations for the SNAIL-TWPA system, where each unit cell consists of a SNAIL shunted by a capacitance $C_g$, and $C_J$ denotes the Josephson capacitance of the SNAIL.
The Josephson phase difference of the $n$-th SNAIL satisfies the following 
circuit equation:
\al{
    &\frac{\dd^2\phi_n}{\dd t^2}
        -r\frac{\dd^2}{\dd t^2}\left(\phi_{n+1}-2\phi_n+\phi_{n-1}\right)
        -\omega_0^2\sum_{j=1}^3
            \biggl[
                \frac{c_{j+1}}{j!}\left(\phi_{n+1}^j-2\phi_n^j+\phi_{n-1}^j\right)
            \biggr]
    =0
    \,,
}
where $r=C_J/C_g$ and $\omega_0=1/\sqrt{L_0C_g}$, with $L_0$
being the effective linear inductance of the SNAIL.
Here, the last term originates from the current–phase response characteristic of the SNAIL element, 
where $c_2 = 1$ and $c_3$, $c_4$ are the magnetic-flux-dependent nonlinear coefficients. 
This tunability enables one to control which nonlinear coefficient, $c_3$ or $c_4$, dominates without modifying the circuit hardware.
In the continuum approximation, the circuit equation becomes~\cite{Ranadive:2021fuo}
\al{
	\frac{\pd^2\phi}{\pd t^2}-ra^2\frac{\pd^4\phi}{\pd x^2\pd t^2}
		-v_0^2\frac{\pd^2}{\pd x^2}\left(\phi +\frac{c_3}{2!}\phi^2+\frac{c_4}{3!}\phi^3\right)=0
	\,,\label{eq:circuit eq}
}
where $v_0=a\omega_0$ with unit cell length $a$.
We now divide the phase difference $\phi$ into two pieces $\phi =\overline\phi+\delta\phi$: 
the background solution
$\overline\phi$ which describes the background soliton, 
and the weak probe signal $\delta\phi$
living on top of the background soliton field.

We then derive classical background wave solutions that propagate without changing their shape, 
namely solitons.
We use the reductive perturbation method to derive the scale-invariant 
nonlinear evolution equation 
admitting the stationary wave solution.
To do so, we employ the stretched variables through the Gardner-Morikawa 
transformation defined as
\al{
    \xi =\epsilon^{1/2}(x-v_0t)\,,\ \ 
    \tau =\epsilon^{3/2}t\,,
    \label{eq:GM transformation}
}
where we have introduced the perturbation parameter $\epsilon$ to keep track of the orders in the expansion.
In addition to the above scaling, we consider the expansion of $\phi$ with respect to $\epsilon$ as $\overline\phi =\epsilon^i\overline\phi^{(1)}+\epsilon^{2i}\overline\phi^{(2)}+\cdots$\,, where $i$ should be determined by requiring the balance between the dispersion and nonlinear effects.
Since one can show that the leading-order equation provides
a trivial solution, we will consider the first non-vanishing
equation hereafter.

For the case of $c_3\neq 0$ and $c_4=0$, 
and setting $i=1$, we then extract the ${\cal O}(\epsilon^3)$ terms from Eq.~\eqref{eq:circuit eq} to obtain
\al{
	\frac{\pd \overline\phi^{(1)}}{\pd\tau}+\frac{1}{2}ra^2v_0\frac{\pd^3 \overline\phi^{(1)}}{\pd\xi^3}
	+\frac{1}{2}c_3v_0\overline\phi^{(1)}\frac{\pd\overline\phi^{(1)}}{\pd\xi}
		=0
	\,.\label{eq:e^3 eq}
}
One finds that this equation coincides with the Korteweg-de Vries (KdV) equation~\cite{Korteweg:1895} and is known to have a soliton solution~\cite{Kivshar:1989ue}.
A single soliton solution is given by
\al{
    \overline\phi_{\rm KdV}^{(1)}
        =A\,{\rm sech}^2\left(2(\xi -\beta v_0\tau)/w\right)
    \,,
}
where $A$, $w=2a\sqrt{12r/c_3A}$, and $\beta=c_3A/6$ denote the amplitude, 
half-width, and normalized velocity of the soliton with respect to the $(\xi,\tau)$ coordinates.

\begin{figure}[tbp]
    \centering
    \includegraphics[height=60mm]{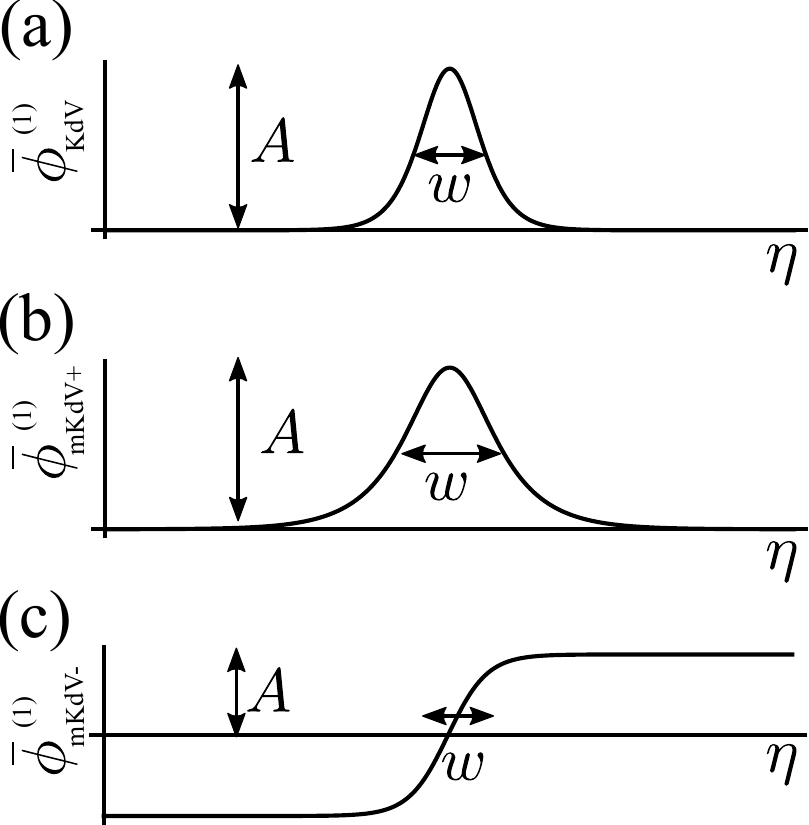}
     \caption{Soliton solutions as a function of $\eta$: (a) KdV soliton, (b) mKdV$^+$ soliton, (c) mKdV$^-$ soliton.
}
     \label{fig:soliton}
\end{figure}

For the opposite case, i.e., $c_3=0$ and $c_4\neq 0$, and setting $i=1/2$, 
we consider the ${\cal O}(\epsilon^{5/2})$ contributions to provide 
the modified KdV equation~\cite{Miura:1968}:
\al{
    \frac{\pd \overline\phi^{(1)}}{\pd\tau}+\frac{1}{2}ra^2v_0\frac{\pd^3 \overline\phi^{(1)}}{\pd\xi^3}
	+\frac{1}{4}c_4v_0\left(\overline\phi^{(1)}\right)^2\frac{\pd\overline\phi^{(1)}}{\pd\xi}
		=0
	\,.\label{eq:e^4 eq}
}
The equation admits different soliton solutions depending on the sign of $c_4$.
For $c_4>0$, this equation admits a single soliton solution given as
\al{
    \overline\phi^{(1)}_{\rm mKdV^+}
        =A\,{\rm sech}\left(2(\xi -\beta v_0\tau) /w\right)
    \,,
}
with $w=2a\sqrt{12r/c_4A^2}$ and $\beta=c_4A^2/24$.
On the other hand, when $c_4<0$, the soliton solution
becomes the shock-wave type~\cite{Perelman:1975,Chanteur:1987}, 
which is given as
\al{
    \overline\phi^{(1)}_{\rm mKdV^-}
        =A\,\tanh\left(2(\xi -\beta v_0\tau) /w\right)
    \,,
}  
with $w=2a\sqrt{12r/|c_4|A^2}$ and $\beta=c_4A^2/12$.
In Fig.~\ref{fig:soliton},
we show the shapes of the soliton solutions as a function of 
$\eta =\xi -\beta v_0\tau$ for 
(a) KdV ($c_3>0$, $c_4=0$),
(b) mKdV$^+$ ($c_3=0$, $c_4>0$),
and (c) mKdV$^-$ ($c_3=0$, $c_4<0$)
models.

\section{Perturbations on soliton background}
\label{sec:3}

In this section, we consider the perturbations on top of 
the soliton background derived in the previous section.
To derive the form of the equation-of-motion for
the weak probe field, it would be convenient to introduce
the coordinate defined as
\al{
    \eta =\xi -\beta v_0\tau
    \,.
}
Since the coordinate $\eta$ can be rewritten in terms of 
the original coordinate $(t,x)$ as
\al{
    \eta =\epsilon^{1/2}\left(x-v_{\rm S}t\right)
    \,,
}
with
\al{
    v_{\rm S}=v_0(1+\beta_{\rm phys})
    \,,
}
the coordinate system $(\eta ,\tau)$ corresponds to
the comoving frame traveling at the soliton velocity $v_{\rm S}$.
Here, we have introduced $\beta_{\rm phys}=\epsilon\beta$ as 
the normalized relative velocity between $v_{\rm S}$ and $v_0$.
We rewrite Eq.~\eqref{eq:circuit eq}
in terms of the comoving coordinate system using the parameters that 
characterize the background soliton.
Moreover, introducing the new variable $\delta\varphi$ defined by $\delta\phi =a\,\pd\delta\varphi/\pd\eta$\,,
integrating the resultant equation along $\eta$ once, and taking the integration constant to be zero,
we obtain the perturbed equation-of-motion for the probe field $\delta\varphi$ as
\al{
	&\epsilon^2\biggl[
		\frac{\pd}{\pd\eta}\left(\frac{v^2-v_{\rm S}^2}{\epsilon}\frac{\pd}{\pd\eta}\right)
		+2v_{\rm S}\frac{\pd^2}{\pd\eta\pd\tau}-\epsilon\frac{\pd^2}{\pd\tau^2}
		+\epsilon ra^2\biggl\{
		\frac{v_{\rm S}^2}{\epsilon}\frac{\pd^4}{\pd\eta^4}-2v_{\rm S}\frac{\pd^4}{\pd\eta^3\pd\tau}
		+\epsilon\frac{\pd^4}{\pd\eta^2\pd\tau^2}\biggr\}
		\biggr]\delta\varphi
		=0
	\,,\label{eq:EoM for deltaphi}
}
where $v(\eta )$ denotes the probe field velocity, which is defined as
\al{
    &v(\eta )=v_0\sqrt{1+c_3\overline\phi(\eta)+\frac{1}{2}c_4\overline\phi{}^2(\eta)}
    \,.
}
In the case of the single soliton solution discussed
in the previous section, the probe field velocity
$v(\eta)$ can be described as
\al{
    \frac{v^2}{v_0^2}=\begin{dcases*}
    1+6\,\beta_{\rm phys}\,{\rm sech}^2(2\eta /w)  & if (a)\,KdV\,, \\
    1+12\,\beta_{\rm phys}\,{\rm sech}^2(2\eta /w) & if (b)\,mKdV$^+$\,, \\
    1+6\,\beta_{\rm phys}\,\tanh^2(2\eta /w)  & if (c)\,mKdV$^-$\,,
    \label{eq:v^2}
  \end{dcases*}
}
where the velocity difference $\beta_{\rm phys}=\epsilon\beta$ 
and half-width of the soliton $w$ were defined for each model in the previous section.

In order to connect the equation for the probe field Eq.~\eqref{eq:EoM for deltaphi} 
to the 
effective metric, we will make several assumptions in the subsequent analysis.
Firstly, we will focus only on the temporal and spatial regions 
where nonlinear dispersion can be neglected.
Specifically,
we will assume that the contributions from the higher-derivative terms 
in Eq.~\eqref{eq:EoM for deltaphi} 
can be neglected. 
This is valid when the timescale of the phenomenon of interest is 
shorter than the timescale at which nonlinear dispersion becomes effective.
We will discuss this issue again in Sec.~\ref{sec:Effect of nonlinear dispersion}.
We further assume that, for the perturbed quantities, 
we do not perform the expansion with respect to 
the parameter of the Gardner-Morikawa transformation
Eq.~\eqref{eq:GM transformation}, namely $\epsilon$.
In other words, we keep the time derivative on the probe field. 
Although the ratio of the length scale to the timescale for the background soliton field is suppressed by the factor $\epsilon$,
for the probe field $\delta\phi$ living on top of the background soliton field, 
the timescale is expected to be determined by $1/\Omega$, where $\Omega$ denotes 
the frequency of the probe field.
Since this may generally differ from that of the background, 
we expect that the time-dependence of the probe field
cannot be generally neglected.
In this paper, we focus on the slowest-decaying mode, 
namely the fundamental mode,
to discuss the stability of the system.
We assume that the frequency $\Omega$ is greater than that derived from the timescale of the background soliton, 
thus the time-derivative terms in Eq.~\eqref{eq:EoM for deltaphi} cannot be neglected.

\begin{figure*}
    \centering
    \includegraphics[height=60mm]{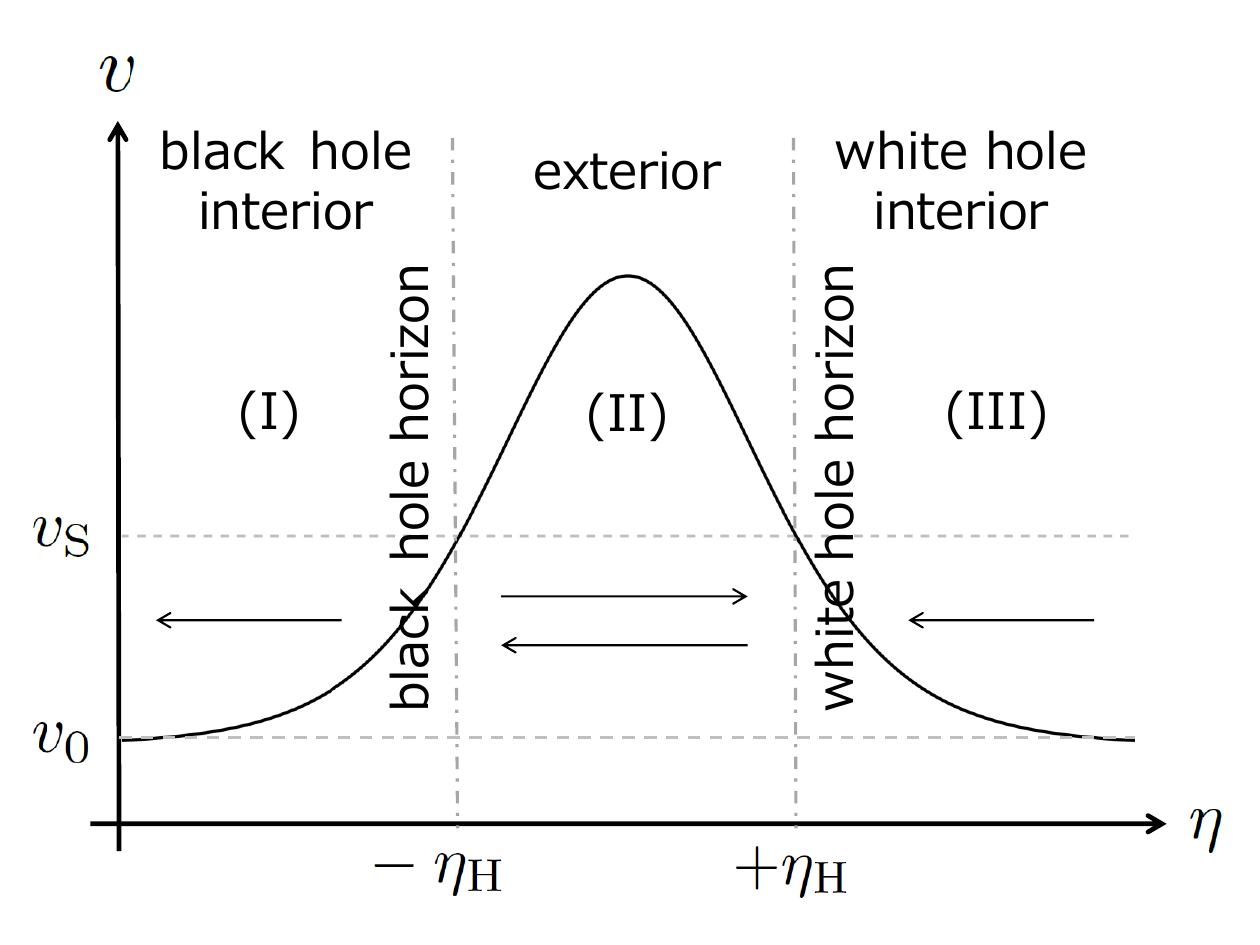}
     \caption{
     The probe field velocity in 
     the background single, right-propagating soliton solution
     as a function of the comoving coordinate 
     $\eta =\epsilon^{1/2}(x-v_{\rm S}t)$.
     Two event horizons are formed at
     $\eta =\pm\eta_{\rm H}$ such that 
     $v^2(\eta =\pm\eta_{\rm H})=v_{\rm S}^2$\,. 
     Region (I) and (III) represent
     the interior regions of black hole
     and white hole, respectively, while
     region (II) corresponds to
     the exterior region of black and white holes. The arrows indicate the characteristic propagation directions of the probe field in the comoving frame.
	}
     \label{fig:configration}
\end{figure*}

With these assumptions, 
keeping the time-derivative term and
dropping the higher-derivative terms in Eq.~\eqref{eq:EoM for deltaphi},
we find
\al{
	\epsilon^2\biggl[
		\frac{\pd}{\pd\eta}\left(\frac{v^2-v_{\rm S}^2}{\epsilon}\frac{\pd}{\pd\eta}\right)
			+2v_{\rm S}\frac{\pd^2}{\pd\eta\pd\tau}-\epsilon\frac{\pd^2}{\pd\tau^2}
		\biggr]\delta\varphi
		=0
	\,.\label{eq:KG eq}
}
This is just the two-dimensional Klein-Gordon equation for the scalar field $\delta\varphi$ 
living on top of the background spacetime described by the effective metric $g_{\mu\nu}$.
The inverse of the effective metric is given by~\cite{Katayama:2022qmr}
\al{
	g^{\mu\nu}
	=\frac{1}{\epsilon\sqrt{-g}}
	\begin{pmatrix} 
 	-\epsilon^2 & \epsilon v_{\rm S}\\
 	\epsilon v_{\rm S}& v^2-v_{\rm S}^2\\
	\end{pmatrix}
	\,,
}
which obviously leads to
\al{
	\dd s^2=&\frac{\sqrt{-g}}{v^2}\Bigl[-\frac{v^2-v_{\rm S}^2}{\epsilon}\dd\tau^2+2v_{\rm S}\dd\tau\dd\eta +\epsilon\dd\eta^2\Bigr]
	\,.
}
To see the causal structure of this
metric clearly, 
we eliminate the cross term between time and space 
and diagonalize the metric by using the coordinate transformation.
Defining the time coordinate as
\al{
	\dd\widetilde\tau =\dd\tau -\frac{\epsilon v_{\rm S}\dd\eta}{v^2-v_{\rm S}^2}
	\,,
}
we obtain the diagonalized metric, 
whose explicit form is given by
\al{
	\dd s^2=\frac{\sqrt{-g}}{\epsilon v^2(\eta )}\Biggl[-\left(1-\frac{v_{\rm S}^2}{v^2(\eta )}\right)v^2(\eta )\dd\widetilde\tau^2 +\frac{\epsilon^2\dd\eta^2}{1-\frac{v_{\rm S}^2}{v^2(\eta )}}\Biggr]
	\,.
}
One finds that the effective metric in this setup 
appears to be conformally related to the two-dimensional part of the Schwarzschild metric,
with the spatially modulated propagation
velocity of the probe field
corresponding to microwave electromagnetic modes.
We note that the spacetime structure of
this analogue system is different 
from that of standard black holes 
that can be found in the universe.
The schematic picture of the
global structure of spacetime 
in our system
is shown in Fig.~\ref{fig:configration}.
For a background configuration corresponding to 
a single, right-propagating 
soliton solution with velocity 
$v_{\rm S}>0$, 
the probe field in the comoving frame 
of a soliton can move in either direction in regions where $v(\eta ) > v_{\rm S}$\,, but in regions where $v(\eta ) < v_{\rm S}$\,, the probe field can move only to the left.
Therefore, 
two event horizons are formed 
at $\eta =\pm\eta_{\rm H}$ such that
$v^2(\eta =\pm\eta_{\rm H})=v_{\rm S}^2$\,.
Region (I) corresponds to the interior of a black hole 
because the probe field moves to the left and cannot reach 
the event horizon. 
In contrast, the probe field in region (III) is always swept 
from the event horizon into region (II). 
Therefore, region (III) corresponds to the interior of a white hole.
In this paper, we focus only on the region $v^2(\eta )\geq v^2_{\rm S}$ 
which corresponds to the exterior
of the black hole/white hole.

\begin{figure*}
    \centering
    \includegraphics[height=60mm]{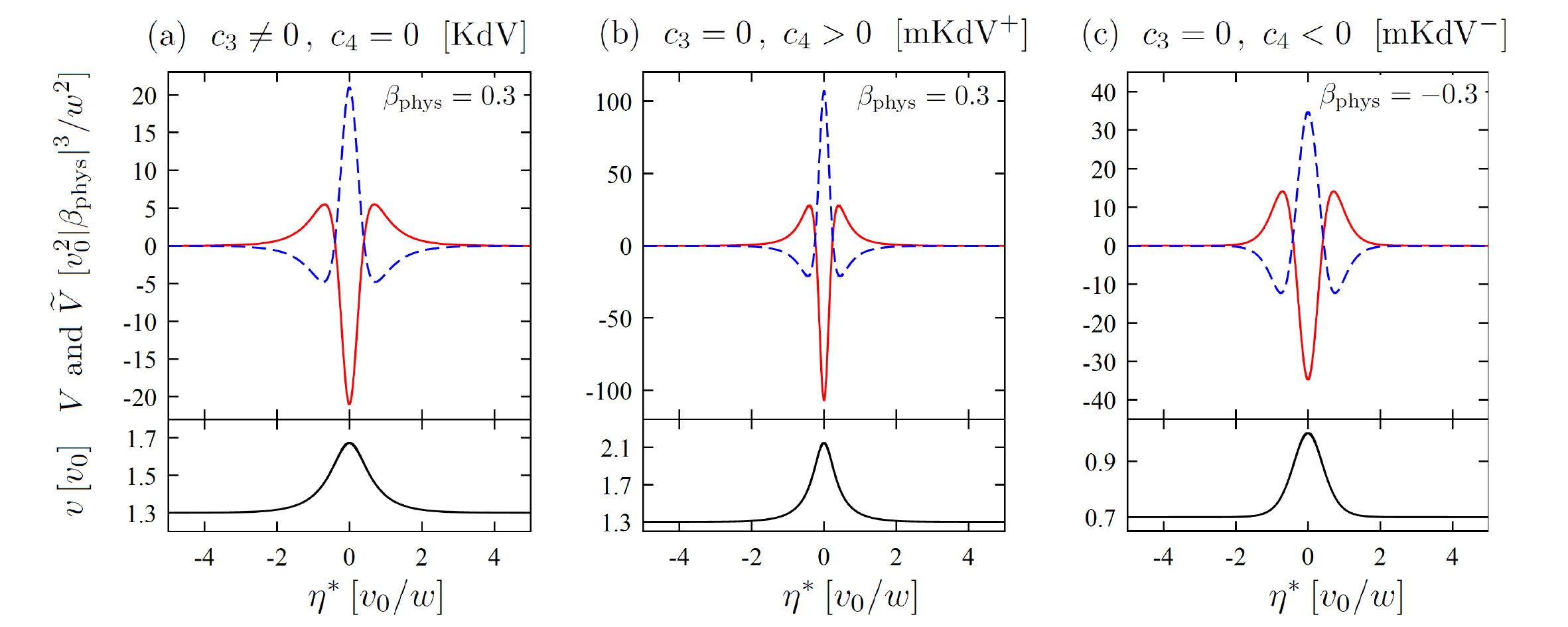}
     \caption{
     Effective potential $V$ (red solid) 
     and its supersymmetric partner potential $\widetilde V$ (blue dashed)
     as functions of the tortoise coordinate $\eta^\ast$ for KdV ($\beta_{\rm phys} =0.3$) [left panel], 
     mKdV$^+$ ($\beta_{\rm phys} =0.3$) [center panel], and mKdV$^-$ ($\beta_{\rm phys} =-0.3$) [right panel] models.
     We also plot the probe field velocity $v$.
	}
     \label{fig:Veff}
\end{figure*}

We then rewrite the Klein-Gordon equation \eqref{eq:KG eq} in terms of 
the $(\widetilde\tau,\eta)$ coordinate system as
\al{
	\biggl[
        \frac{\pd}{\pd\eta}\left(\frac{v^2-v_{\rm S}^2}{\epsilon}\frac{\pd}{\pd\eta}\right)
        -\frac{\epsilon v^2}{v^2-v_{\rm S}^2}\frac{\pd^2}{\pd\widetilde\tau^2}
    \biggr]\delta\varphi 
    =0
	\,.\label{eq:KG eq2}
}
It can be seen from this equation that in the coordinate system where the metric is diagonalized, 
the derivatives of $\eta$ and $\widetilde\tau$ on the probe field 
provide contributions of the same order since $v^2-v_{\rm S}^2={\cal O}(\epsilon )$.
Performing the Fourier transformation of the perturbation $\delta\varphi$ in terms 
of the $\widetilde\tau$ coordinate, which is defined as
\al{
	&\delta\varphi (\widetilde\tau,\eta )=\int\frac{\dd\Omega}{2\pi}\,\delta\varphi (\Omega, \eta)e^{-{\rm i}\Omega\widetilde\tau}
	\,.\label{eq:Fourier}
}
The Fourier counterpart of the Klein-Gordon equation \eqref{eq:KG eq2} is
\al{
	\biggl[\frac{v^2-v_{\rm S}^2}{\epsilon v^2}\frac{\dd}{\dd\eta}\left(\frac{v^2-v_{\rm S}^2}{\epsilon}\frac{\dd}{\dd\eta}\right)+\Omega^2\biggr] \delta\varphi =0
	\,.\label{eq:Fourier KG}
}
We then introduce the tortoise coordinate $\eta^\ast$, which is defined by
\al{
	\frac{\dd\eta^\ast}{\dd\eta}=\frac{v}{v^2-v_{\rm S}^2}
	\,.\label{eq:tortoise coordinate}
}
The explicit form of the tortoise coordinate
in terms of the $\eta$ coordinate for each model discussed in
the previous subsection is shown in
Appendix \ref{sec:Tortoise coordinate}.
With the tortoise coordinate, 
we can rewrite Eq.~\eqref{eq:Fourier KG} as
\al{
	\delta\varphi^{\prime\prime}+\frac{v^\prime}{v}\delta\varphi^{\prime}+\epsilon^2\Omega^2\delta\varphi=0
	\,,
}
where the prime denotes the derivative with respect to $\eta^\ast$.
To transform the above Klein-Gordon equation into the Schrödinger-type equation, 
we need to eliminate the first-derivative term.
Setting
\al{
	\delta\varphi (\eta^\ast)=Z(\eta^\ast)H(\eta^\ast)
	\,,
}
we have
\al{
	&H^{\prime\prime}
        +\left(\frac{v^\prime}{v}+2\frac{Z^\prime}{Z}\right) H^\prime
		+\left(\epsilon^2\Omega^2+\frac{v^\prime}{v}\frac{Z^\prime}{Z}+\frac{Z^{\prime\prime}}{Z}\right) H=0
	\,.\label{eq:H eq}
}
We impose the coefficient of $H^\prime$ to be zero:
\al{
	Z\propto\frac{1}{\sqrt{v}}
	\,.
}
Substituting this expression into Eq.~\eqref{eq:H eq}, we finally obtain the Schrödinger-type equation for $H$ as
\al{
	-H^{\prime\prime}+V(\eta^\ast ) H=\epsilon^2\Omega^2 H
	\,,\label{eq:H eq2}
}
where the effective potential is given by
\al{
	V=\frac{1}{2}\frac{v^{\prime\prime}}{v}
        -\frac{1}{4}\left(\frac{v^\prime}{v}\right)^2
		=\frac{1}{\sqrt{v}}\left(\sqrt{v}\right)^{\prime\prime}
	\,,\label{eq:V}
}
which implies
that $V$ approaches zero near the event horizon, namely $v\to v_{\rm S}$, since 
$v^\prime\propto v^2-v_{\rm S}^2$\,.
Therefore, solving the equation \eqref{eq:H eq2} can be treated as a scattering problem
in standard quantum mechanics.
We plot the effective potential $V$ and probe field velocity $v$ as functions of
the tortoise coordinate for (a) KdV ($c_3\neq 0\,,c_4=0$) with $\beta_{\rm phys} =0.3$, 
(b) mKdV$^+$ ($c_3=0\,, c_4>0$) with $\beta_{\rm phys}=0.3$, and (c) mKdV$^-$ ($c_3=0\,,c_4<0$) 
with $\beta_{\rm phys} =-0.3$ in Fig.~\ref{fig:Veff}.
We find that all the effective potentials share a volcano-type shape and exhibit a negative dip at $\eta^\ast = 0$.

\section{Linear stability and supersymmetric partner}
\label{sec:4}

In this section, we discuss the linear stability of the system by using
the information regarding the potential shape.
When we would like to prove the stability of 
spacetime, we need 
to show the non-existence of normalizable $\Omega^2<0$ solutions,
namely, the exponentially growing mode, under the appropriate (vanishing) boundary conditions, 
$H\to 0$\,, $H^\prime\to 0$
at $\eta^\ast\to\pm\infty$ and the conditions such that $H$ and 
$H^\prime$ are continuous and bounded
everywhere.
Multiplying the complex conjugate of $H$, say $\overline H$, 
and integrating $\eta^\ast$ in Eq.~\eqref{eq:H eq2}, we have
\al{
	&-\Bigl[\overline HH^\prime\Bigr]_{-\infty}^\infty
		+\int_{-\infty}^\infty\dd\eta^\ast
        \Bigl(\bigl|H^\prime\bigl|^2+V\bigl|H\bigl|^2\Bigr)
        =\epsilon^2\Omega^2\int_{-\infty}^\infty\dd\eta^\ast\bigl|H\bigl|^2
	\,.
}
When one considers that the effective potential $V$ is non-negative everywhere,
one finds that $\Omega^2$ is manifestly non-negative.
Since, as we have already pointed out in the previous section, 
the effective potential $V$ in our case contains the negative region, 
we cannot apply the above formalism directly.
However, we still demonstrate the stability of our spacetime 
against perturbations in the following way (see, e.g., \cite{Kimura:2017uor}). 
Let us introduce the supersymmetry generators as
\al{
	\widehat Q=\frac{\dd}{\dd\eta^\ast }+W(\eta^\ast )
	\,,\ \ \ \ 
	\widehat Q^\dagger =-\frac{\dd}{\dd\eta^\ast}+W(\eta^\ast )
	\,.
}
When the superpotential $W$ is chosen as $W=v^\prime /2v$, it can be shown that
the equation-of-motion for $H$ 
can be rewritten in terms of the supersymmetric generators, namely
\al{
	\widehat Q\widehat Q^\dagger H
	=\epsilon^2\Omega^2 H
	\,.\label{eq:SUSY QM}
}
This is known as the supersymmetric quantum mechanics system~\cite{Cooper:1994eh}.
With this expression, multiplying $\overline H$ and integrating $\eta^\ast$ in Eq.~\eqref{eq:SUSY QM}, we find
\al{
	\int_{-\infty}^\infty\dd\eta^\ast 
		\overline H\widehat Q\widehat Q^\dagger H
        =&\Bigl[\overline H\widehat Q^\dagger H\Bigr]_{-\infty}^\infty
		+\int_{-\infty}^\infty\dd\eta^\ast\bigl|\widehat Q^\dagger H\bigl|^2
		=\epsilon^2\Omega^2\int_{-\infty}^\infty\dd\eta^\ast\bigl|H\bigl|^2
	\,.
}
When we impose the additional boundary condition such that $W$ is not divergent at $\eta^\ast\to\pm\infty$, 
the boundary term vanishes, which implies that $\Omega^2 >0$, namely
the non-existence of the exponentially growing (instability) mode.
Since $W\propto v^\prime/v\propto (v^2-v_{\rm S}^2)$ 
in our soliton background, the boundary condition of $W$ is
naturally satisfied.
Therefore, we conclude that our analogue system is perturbatively stable
in regions where the system can be described by the effective theory in which nonlinear dispersion is negligible.
\\

In the context of supersymmetric quantum mechanics, it would be useful to consider the supersymmetric partner, which is defined as
\al{
	&\widetilde H
		=\frac{1}{\Omega}\,\widehat Q^\dagger H
	\,.
}
If the state $H$ is an eigenfunction of $\widehat Q\widehat Q^\dagger$, then 
its supersymmetric partner $\widetilde H$ is an eigenfunction of 
$\widehat Q^\dagger \widehat Q$ with the same eigenvalue.
In particular, the equation-of-motion for $\widetilde H$ represents
\al{
	\widehat Q^\dagger \widehat Q\,\widetilde H
		=-\widetilde H^{\prime\prime}+\widetilde V(\eta^\ast )\widetilde H
		=\epsilon^2\Omega^2\widetilde H
	\,,\label{eq:SUSY EoM}
}
with
\al{
	\widetilde V=-W^\prime+W^2
    =-\frac{1}{2}\frac{v^{\prime\prime}}{v}
        +\frac{3}{4}\left(\frac{v^\prime}{v}\right)^2
	\,.
}
In Fig.~\ref{fig:Veff}, we plot the effective potential 
for the supersymmetric partner in addition to that for the original state.
Since the effective potential of the original state $V$ 
is concave downward,
it is not straightforward to solve the scattering problem in order 
to determine the QNM frequency. 
Unlike the effective potential of the original state,
the effective potential for the supersymmetric partner $\widetilde V$ possesses
a positive value at its peak position and is convex upward. 
Since it has an upward-convex structure at the peak, we expect that existing methods for
calculating the QNM, particularly the WKB method, can be applied with minimal modifications.
Therefore, we shall extract the information regarding 
the value of the QNM frequency by solving the equation 
for the supersymmetric partner possessing
the same energy eigenvalues 
(see also \cite{Chandrasekhar:1975nkd}). 

We have seen above that $H$ and $\widetilde H$ share the eigenvalue
$\epsilon^2\Omega^2$.
For the two problems to give the same QNM frequencies, we also need the outgoing
boundary conditions to be preserved, and the correspondence between $H$ and
$\widetilde H$ to be invertible.
Both follow from the behaviour of $W$ noted above:
$W=v^\prime /2v$ is finite throughout the region of interest, since
$v^2\geq v_{\rm S}^2>0$ there, and $W\propto v^2-v_{\rm S}^2$ vanishes at the two
horizons.
The relation $\widetilde H=\widehat Q^\dagger H/\Omega$ hence maps an outgoing
solution to an outgoing solution of non-vanishing amplitude, and so does the
inverse relation $H=\widehat Q\widetilde H/(\epsilon^2\Omega)$.
The correspondence breaks down only at $\Omega =0$, where $\widehat Q^\dagger H$
vanishes for $H\propto\sqrt{v}$, that is for $\delta\varphi ={\rm const.}$, which
is a trivial solution and is discarded.
No other frequency is exceptional in this respect, since $W$ does not depend on
$\Omega$ in our system.
The QNM frequencies obtained from $\widetilde V$ therefore agree with those of
$V$ for every $\Omega\neq 0$, as we have also confirmed numerically.

\section{Quasi-normal mode frequency in analogue system}
\label{sec:5}

The QNM in our analogue system can be defined in the usual way 
by imposing appropriate boundary conditions and solving
the corresponding eigenvalue problem.
Close to the event horizons of 
the black hole and white hole, we impose
the outgoing boundary conditions as
\al{
	&H\propto e^{\pm{\rm i}\epsilon\Omega\eta^\ast}\ \ (\eta^\ast\to\pm\infty)
	\,.\label{eq:bc}
}
At the black-hole horizon, the mode falls into the black-hole interior, whereas at the white-hole horizon, it approaches the horizon from region (II) but cannot cross into the white-hole interior; the mode emerging from the white-hole interior is excluded.
With these boundary conditions, we would like to solve the Schrödinger-type equation 
\eqref{eq:H eq2} to obtain
the eigenvalue, which corresponds to the square of the QNM frequency.

In the case of a realistic black hole solution, the functional forms of the tortoise coordinate 
and the effective potential are known analytically.
However, in the case of our analogue system, the functional form of 
the effective potential as a function of the tortoise coordinate can be determined numerically
because the effective potential depends on
the probe field velocity $v$ and its derivatives with respect to the tortoise coordinate,
whose definition depends on $v$ itself.
Therefore, numerical methods must be used to determine the QNM frequency.
We adopt the shooting method in the subsequent analysis.
As a reference for investigating parameter dependencies, we will also conduct a comparison with methods that use the WKB approximation~\cite{Schutz:1985km,Iyer:1986np,Iyer:1986nq,Konoplya:2003ii,Matyjasek:2017psv,Konoplya:2019hlu}.

\subsection{Shooting method}
\label{sec:shooting method}

In the shooting method, 
we integrate the perturbation equation from one end of the numerical domain to the other end and fine-tune the initial condition to find the solution satisfying the correct boundary condition at the other end.
For numerical convenience 
in the subsequent analysis, 
we will solve the original wave equation~\eqref{eq:Fourier KG}
using the $\eta$ coordinate instead of 
the tortoise $\eta^\ast$ coordinate.

The procedure of the shooting method is as follows
(see Appendix \ref{sec:Details of the shooting method} for more details).
We take the region $\eta \in (-\eta_{\rm H}, \eta_{\rm H})$ between the two horizons at $\eta = \pm\eta_{\rm H}$ as the numerical domain, where we have taken the convention $\eta_{\rm H} > 0$. 
From the two horizons at $\eta = \pm \eta_{\rm H}$ toward some intermediate value of $\eta =\eta_{\rm c}$\,, 
we integrate Eq.~\eqref{eq:Fourier KG} to construct two solutions $\delta\varphi_\pm (\eta)$ satisfying the boundary conditions at $\eta = \pm \eta_{\rm H}$, respectively, for a given $\Omega$\,.
In the $\eta$ coordinate, the QNM boundary condition is given by
\al{
    \delta\varphi_\pm (\eta )\propto
    \bigl|\eta \mp \eta_{\rm H}\bigr|^{-{\rm i}\epsilon\Omega /\kappa} \ \ (\eta\to \pm \eta_{\rm H} )
    \,,
    \label{eq:BC_eta}
}
where $\kappa$ is a constant depending on the parameters $v_0$, $w$ and $\beta_{\rm phys}$ (see Eq.~\eqref{eq:kappa-def} for the explicit form of $\kappa$ in each model).
For $\delta\varphi_\pm$ to represent a correct mode function satisfying all the boundary conditions,
the two solutions $\delta\varphi_\pm$ must be 
linearly dependent at the matching point $\eta =\eta_{\rm c}$.
It implies that the Wronskian of the two solutions must vanish, namely
\al{
    W(\Omega )
    =\delta\varphi_+ (\eta_{\rm c})\dot{\delta\varphi}_-(\eta_{\rm c})-\dot{\delta\varphi}_+ (\eta_{\rm c})\delta\varphi_-(\eta_{\rm c})
    =0
    \,,\label{eq:Wronskian}
}
which gives the equation for the QNM frequency $\Omega$ (a dot denotes a derivative with respect to $\eta$).
Furthermore, we can simplify the problem
by utilizing the symmetry of our system.
Since the wave equation~\eqref{eq:Fourier KG} and also the boundary condition~\eqref{eq:BC_eta} are symmetric under the transformation $\eta \to -\eta$,
the two solutions $\delta\varphi_\pm$ are related to each other as $\delta\varphi_+(\eta)=C\, \delta\varphi_-(-\eta)$\,, where $C$ is a proportional constant.
It implies that
the Wronskian Eq.~\eqref{eq:Wronskian} for $\eta_{\rm c}=0$
can be reduced to 
\al{
    W(\Omega) \propto \delta\varphi_-(0)\dot{\delta\varphi}_- (0)=0
    \,.
    \label{eq:Wronskian-reduced}
}
Therefore, we found that it is sufficient to
search for $\Omega$ satisfying either
$\delta\varphi_-(0)=0$ or $\dot{\delta\varphi}_- (0)=0$\,
by some root-finding algorithm.

\begin{figure*}
    \centering
    \includegraphics[height=45mm]{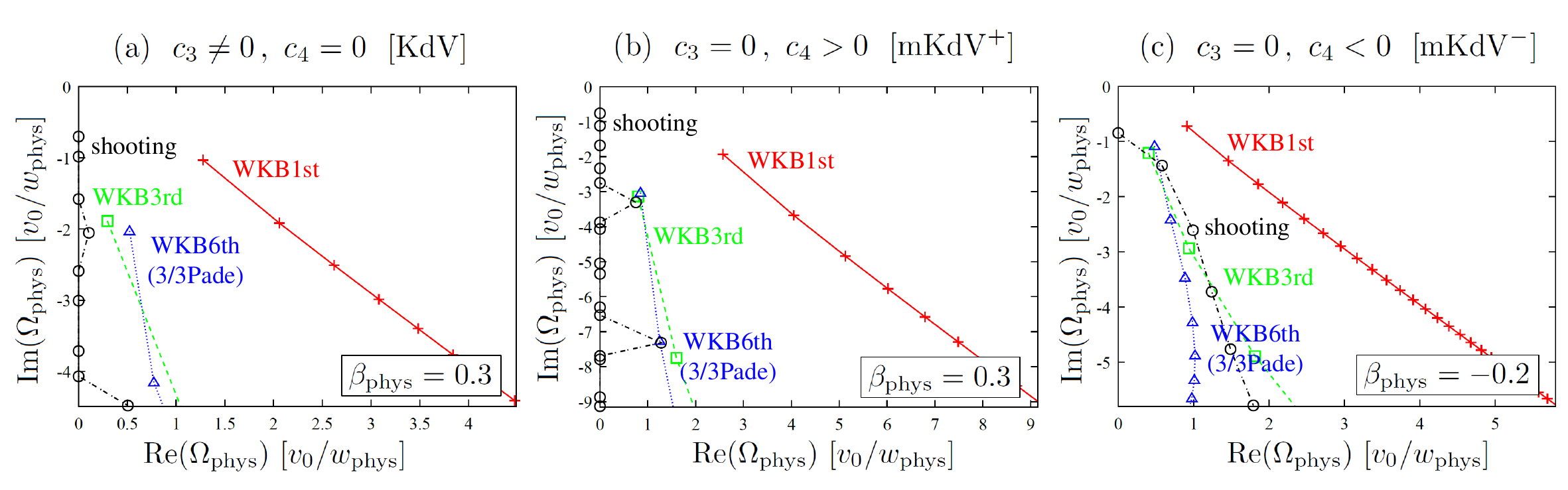}
     \caption{
    Quasi-normal mode spectra for analogue black-white holes constructed by 
    SNAIL-TWPA circuit system 
    evaluated by the shooting method
    are shown in black $\bigcirc$.
    For comparison, the quasi-normal mode spectra
    with various WKB orders;
    the first order [red $+$], the third order [green $\Box$], 
    and the sixth order
    with the 3/3 Padé approximation [blue $\bigtriangleup$].
	}
     \label{fig:QNM_WKBorder}
\end{figure*}

We plot in Fig.~\ref{fig:QNM_WKBorder} the QNM frequencies obtained
by the shooting method for the case of $\beta_{\rm phys}=0.3$ in
(a) KdV ($c_3\neq 0$, $c_4=0$) and (b) mKdV$^+$ ($c_3=0$, $c_4>0$), and $\beta_{\rm phys}=-0.2$ in (c) mKdV$^-$ ($c_3=0$, $c_4<0$).
Table~\ref{table:QNM} also shows the frequency of 
the least-damped QNM,
namely the fundamental mode, evaluated using
the shooting method for different soliton models.
To examine the $\beta_{\rm phys}$ dependence,
we show in Table \ref{table:QNM}
the results for $\beta_{\rm phys}=0.3$\,, $0.5$\,,
$0.8$ in (a) KdV and (b) mKdV$^+$\,, and
for $\beta_{\rm phys}=-0.1$\,, $-0.15$\,, $-0.2$ 
in (c) mKdV$^-$.
The numerical data summarized in Table~\ref{table:QNM} indicates $|\Omega_\text{phys}| \propto |\beta_\text{phys}|{}^p$ with $p\sim 0.7$.

\begin{table*}[tb]
\centering
  \caption{Frequencies of the least-damped quasi-normal modes 
  for different background soliton models, 
  evaluated with the shooting method (shooting).
  For reference, the results from 
  the 6th-order WKB method with the 3/3 Padé approximation (WKB6th) are also shown.}
  \label{table:QNM}
  \vspace{1mm}
\begin{tabular}{|l|c|c|c|c|c|c|}  \hline
    \multicolumn{7}{|c|}{$\Omega_{{\rm phys},n}\, [v_0/w_{\rm phys}]$ } \\
    \hline\hline
     & \multicolumn{2}{|c|}{$\beta_{\rm phys} =0.3$} & \multicolumn{2}{|c|}{$\beta_{\rm phys} =0.5$} & \multicolumn{2}{|c|}{$\beta_{\rm phys}=0.8$} \\ 
   \hline
   model & shooting & WKB6th & shooting  & WKB6th  & shooting  & WKB6th \\ 
   \hline
    (a) KdV &\  $0.00-0.70{\rm i}$\ \   &\  ($0.52-2.03{\rm i})$\ \  &\  $ 0.00 - 1.01 {\rm i}$\ \  &\  ($0.63-2.49{\rm i}$)\ \  &\  $ 0.00 - 1.38 {\rm i}$\ \ &\  ($0.76-2.94{\rm i}$)\ \  
   \\
   (b) mKdV$^+$ &\  $0.00-0.76{\rm i}$\ \ &\ ($0.85-3.03{\rm i})$\ \  &\ $ 0.00 - 1.07 {\rm i}$\ \ &\ ($1.12-3.72{\rm i}$)\ \ &\ $ 0.00 - 1.45 {\rm i}$\ \  &\ ($1.46-4.55{\rm i}$)\ \ 
   \\
   \hline\hline
   & \multicolumn{2}{|c|}{$\beta_{\rm phys} =-0.1$} & \multicolumn{2}{|c|}{$\beta_{\rm phys} =-0.15$} & \multicolumn{2}{|c|}{$\beta_{\rm phys}=-0.2$} \\ 
   \hline
   (c) mKdV$^-$ &\ $ 0.00 - 0.45 {\rm i}$\ \ & \ ($0.09-0.79{\rm i}$)\ \ &\ $ 0.00 - 0.66 {\rm i}$\ \ &\ ($0.23-0.98{\rm i}$)\ \ &\ $0.00-0.84 {\rm i}$\ \  &\ ($0.48-1.08{\rm i}$)\ \ 
   \\
   \hline
 \end{tabular}
\end{table*}

\subsection{WKB method}

We employ the WKB method to estimate the analytical structure of the QNM frequency~\cite{Schutz:1985km,Iyer:1986np,Iyer:1986nq,Konoplya:2003ii,Matyjasek:2017psv,Konoplya:2019hlu}.
Since, as already mentioned in the previous section, 
the effective potential for the original state $H$ in our system contains 
the negative region and is convex upwards, 
this differs from the case of the QNMs in the context of the standard black hole perturbations.
Therefore, we shall cease solving the QNM frequency for the original state and instead solve the equation for its supersymmetric partner $\widetilde H$ possessing the same eigenvalue, Eq.~\eqref{eq:SUSY EoM}.
Eq.~\eqref{eq:bc} corresponds to the boundary condition of the supersymmetric partner 
$\widetilde H$ as
\al{
	\widetilde H\propto e^{\pm{\rm i}\epsilon\Omega\eta^\ast}\ \ (\eta^\ast\to\pm\infty)
	\,.
    \label{eq:H_BC}
}
Hence, we can impose the same boundary condition as that of $H$, 
since $v^{\prime}\propto v^2-v_{\rm S}^2$ approaches zero near the boundaries.
As shown at the end of Sec.~\ref{sec:4}, the QNM spectra of $V$ and
$\widetilde V$ coincide for all $\Omega\neq 0$, so that the WKB analysis of
$\widetilde V$ below determines the QNM frequencies of the original problem
as well.

The WKB approximation method constructs the QNMs by approximating $\widetilde H$
with WKB functions on both sides of the potential barrier, 
matching across the potential peak
and imposing the outgoing boundary conditions on it.
The WKB formula for the QNM frequency is given by
\al{
	&\left(\epsilon\Omega_n^{\rm WKB}\right)^2=\widetilde V_0-{\rm i}\left(n+\frac{1}{2}\right)\sqrt{-2\widetilde V^{\prime\prime}_0}+\cdots
	\,,\label{eq:omega2}
}
where $n$ is a non-negative integer and
the subscript $0$ denotes the evaluation at $\eta^\ast =\eta^\ast_0$ 
with $\eta^\ast_0$ being the position of the potential peak.
Since the frequency corresponding to the physical time $t$ is determined
as $\Omega_{{\rm phys},n}=\epsilon^{3/2}\Omega_n$, we have
\al{
    \left(\Omega_{{\rm phys},n}^{\rm WKB}\right)^2
        =\epsilon\biggl[
            \widetilde V_0-
            {\rm i}\left(n+\frac{1}{2}\right)\sqrt{-2\widetilde V^{\prime\prime}_0}+\cdots
            \biggr]
    \,.
    \label{eq:omega2_phys}
}
Although the explicit expressions of the higher-order corrections 
are not shown here, in the subsequent analysis, we will use
the WKB formula valid up to the sixth WKB order~\cite{Konoplya:2003ii,Konoplya:2019hlu}.
We also use the 3/3 Padé approximation of the sixth-order WKB formula. 
We note that in our analogue system, the correction terms of the WKB approximation
are not small compared with the leading term for any value of $\beta_{\rm phys}$,
and they grow further when $|\beta_{\rm phys}|$ is sufficiently small,
say $|\beta_{\rm phys}|\lesssim 0.1$ (see Appendix \ref{sec:Validity of WKB approximation}).
Therefore, in this paper, we adopt values of $\beta_{\rm phys}$ for which
the correction terms are comparatively small, and present the WKB results for reference.
As a demonstration, we show the results for three cases: 
(a) KdV 
with $\beta_{\rm phys} =0.3$, 
(b) mKdV$^+$ 
with $\beta_{\rm phys}=0.3$, 
and (c) mKdV$^-$  
with $\beta_{\rm phys} =-0.2$ in Fig.~\ref{fig:QNM_WKBorder}.
We adopt the first-order [red $+$], third-order [green $\Box$],
and sixth-order 3/3 Padé [blue $\bigtriangleup$] formula of the WKB method.
This figure shows that the resultant frequency 
of the fundamental mode is
of the same order of magnitude in all WKB orders.
In particular, the third- and sixth-order results
of the fundamental frequency yield
nearly identical values.
We also show in Table~\ref{table:QNM} the QNM fundamental mode 
evaluated by the sixth-order 3/3 Padé 
for $\beta_{\rm phys}=0.3$\,, $0.5$\,,
$0.8$ in (a) KdV and (b) mKdV$^+$\,, and
for $\beta_{\rm phys}=-0.1$\,, $-0.15$\,, $-0.2$ 
in (c) mKdV$^-$.

In order to derive the simple expression for the parameter dependence
of the QNM frequency, we consider the lowest order expansion 
when $\beta_{\rm phys}$ is small for simplicity in analysis.
Ignoring the numerical coefficients and higher-order correction terms,
we show 
the square of the leading-order QNM frequency as
${\rm Re}([\Omega_{{\rm phys},n}^{\rm WKB\,1st}]^2)\approx\epsilon\widetilde V_0\propto\beta_{\rm phys}^3v_0^2/w_{\rm phys}^2\,,$
and ${\rm Im}([\Omega_{{\rm phys},n}^{\rm WKB\,1st}]^2)\approx\epsilon\sqrt{-\widetilde V^{\prime\prime}_0}\propto\beta_{\rm phys}^{5/2}v_0^2/w_{\rm phys}^2$\,,
where $w_{\rm phys}=w/\sqrt{\epsilon}$ denotes the physical half-width of the soliton.
The real part of the square of the QNM frequency 
is suppressed by the factor $\beta_{\rm phys}^{1/2}$ and
the imaginary part provides the dominant contributions to 
$\Omega_{{\rm phys},n}^2$.
Hence, we obtain the approximate form: $\Omega_{{\rm phys},n}^{\rm WKB 1st}\propto\beta_{\rm phys}^{5/4}v_0/w_{\rm phys}(1-{\rm i})$\,. Based on this, we expect that
the fundamental mode 
of the QNM in our black-white holes
generally exhibits
the following parameter dependence:
\al{
	\Omega_{{\rm phys},0}
		=F(\beta_{\rm phys})\times
        \frac{v_0}{w_{\rm phys}}
	\,.\label{eq:QNM result}
}
The QNM fundamental frequency 
can typically be evaluated as a function of $\beta_{\rm phys}$ 
multiplied by the inverse of the timescale at which the probe field 
reaches the event horizons.
While $F\propto\beta_{\rm phys}^{5/4}$ in the first-order WKB approximation, the functional form of $F$ is slightly modified when the higher-order WKB corrections are taken into account.
As shown in Table~\ref{table:QNM},
the $\beta_{\rm phys}$-dependence of $F$ is
found to be weaker than that observed in the first-order WKB
approximation. Nevertheless, the order of magnitude of the results 
is not expected to change significantly,
although further analysis may reveal minor adjustments.

\subsection{Comparison of shooting and WKB methods}

By comparing the results obtained from the shooting and WKB methods, 
we find both similarities and differences between them.
Although the $\beta_{\rm phys}$-dependence of the fundamental mode estimated from the shooting method is weaker than that from the first-order WKB approximation, 
it is still monotonically increasing with respect to $\beta_{\rm phys}$.
We also find numerical evidence that the fundamental modes are pure imaginary, unlike the WKB method results.

One possible cause of the discrepancy 
is that we underestimated the coefficients 
when evaluating the QNM using a WKB formula 
appropriate for a single-peak potential.
Based on Fig.~\ref{fig:Veff}, we expected 
the potential to be well approximated
by a single-peaked potential; therefore, 
we concluded that the WKB approximation 
could be applied. 
However, contrary to expectations, the correction terms $\Lambda_i$ evaluated 
in App.~\ref{sec:Validity of WKB approximation} are not small compared with the 
leading term $n+1/2$ of the fundamental mode, so that the WKB series is not well 
controlled in our system.
In contrast, we have carefully 
verified the validity of the shooting method, shown in 
App.~\ref{sec:Details of the shooting method}, and therefore believe 
that the results obtained using this method are more reliable than those obtained 
using the WKB method. Nevertheless, we still believe that the quasi-analytic formula 
derived from the WKB method is useful for analysing the dependence of the QNM frequency 
on parameters.

While the validity of the WKB approximation
cannot be guaranteed in our system, the fundamental frequencies obtained from both methods generally have the same order of magnitude.
We also find that the QNM frequencies with nonzero real parts are reproduced to some extent by the WKB results, as we can observe in Fig.~\ref{fig:QNM_WKBorder}. It may indicate that the WKB approximation correctly captures some aspects of the QNM frequencies.

\subsection{Effect of nonlinear dispersion}
\label{sec:Effect of nonlinear dispersion}

In this subsection, we estimate the effect
of the nonlinear dispersion terms
that were neglected in previous analyses.
To achieve this, it is necessary to properly evaluate 
the contribution of the higher-derivative terms 
in the full set of equation-of-motion for $\delta\varphi$.
We rewrite Eq.~\eqref{eq:EoM for deltaphi}
in terms of the coordinate $(\widetilde\tau, \eta^\ast)$ 
and perform the Fourier transformation
as
\al{
    \frac{1}{v}\left(v\,\delta\varphi^\prime\right)^\prime +\epsilon^2\Omega^2\,\delta\varphi 
    =\widehat{\cal E}_{\rm NL}\,\delta\varphi
    \,,\label{eq:full eq}
}
where
\al{
    \widehat{\cal E}_{\rm NL}
    =&-\frac{\epsilon ra^2}{v^2-v_{\rm S}^2}
    \Biggl\{
        \frac{v^2-v_{\rm S}^2}{v}\frac{\dd}{\dd\eta^\ast}
        \left(\frac{v}{v^2-v_{\rm S}^2}\frac{\dd}{\dd\eta^\ast}\right)
        -\frac{2{\rm i}\epsilon v_{\rm S}v^\prime\Omega}{v^2-v_{\rm S}^2}
        +\frac{2{\rm i}\epsilon v_{\rm S}\Omega}{v}\frac{\dd}{\dd\eta^\ast}
        -\frac{\epsilon^2 v_{\rm S}^2\Omega^2}{v^2}
    \Biggr\}
     \notag\\
     &\quad\times \frac{v_{\rm S}^2v^2}{(v^2-v_{\rm S}^2)^2}
     \Biggl\{
        \frac{v^2-v_{\rm S}^2}{v}\frac{\dd}{\dd\eta^\ast}\left(\frac{v}{v^2-v_{\rm S}^2}\frac{\dd}{\dd\eta^\ast}\right)
        -\frac{2{\rm i}\epsilon v_{\rm S}v^\prime\Omega}{v^2-v_{\rm S}^2}
        +\frac{2{\rm i}\epsilon v\Omega}{v_{\rm S}}\frac{\dd}{\dd\eta^\ast}
        -\frac{\epsilon^2v^2\Omega^2}{v_{\rm S}^2}
    \Biggr\}
    \,.\label{eq:nonlinear dispersion term}
}
To evaluate each term, 
it is necessary to determine the magnitude of 
the contribution from 
the $\eta^\ast$ derivative in the regions with 
mildly nonlinear dispersion.
We expect that the $\eta^\ast$ derivative term 
and the time derivative term will be balanced even in such a region,
namely
\al{
    \delta\varphi^\prime ={\cal O}\left(\epsilon\Omega\right)
    \,.\label{eq:etaast derivative}
}
The linear dispersion terms, which correspond to
the left-hand-side of Eq.~\eqref{eq:full eq},
include up to the second-order derivatives with respect to $\eta^\ast$, 
implying
\al{
    (\text{LHS of Eq.~}\eqref{eq:full eq})
        ={\cal O}\left(\epsilon^2\Omega^2\right)
    \,.
}
On the other hand, the nonlinear dispersion term,
which is the right-hand-side of Eq.~\eqref{eq:full eq},
includes up to fourth-order derivatives with respect to $\eta^\ast$.
Using Eq.~\eqref{eq:etaast derivative}, we can evaluate 
the contribution from the nonlinear dispersion as
\al{
    \widehat{\cal E}_{\rm NL}\,\delta\varphi
        =&{\cal O}\left(\frac{\epsilon ra^2}{v_0^2}\epsilon^4\Omega^4\right)
    \,.
}
We then substitute the analytical
expression for the QNM frequency Eq.~\eqref{eq:QNM result}
into the above equation to obtain
\al{
     \widehat{\cal E}_{\rm NL}\,\delta\varphi
        \approx\beta_{\rm phys}\,F^2(\beta_{\rm phys})\times {\cal O}\left(\epsilon^2\Omega^2\right)
    \,.\label{eq:E_NL order}
}
Based on our analysis thus far, we expect $F(\beta_{\rm phys})$ 
to be a positive power of $\beta_{\rm phys}$.
Hence, the contribution from the nonlinear dispersion term
is expected to be sufficiently suppressed relative to 
the linear dispersion terms
($\sim {\cal O}\left(\epsilon^2\Omega^2\right)$)
by $\beta_{\rm phys}$.
We have to note that this estimation is only valid in regions 
that are sufficiently distant from the event horizons.
Near the horizons, the contribution of the term $1/(v^2-v_{\rm S}^2)$ 
becomes dominant, as seen from Eq.~\eqref{eq:nonlinear dispersion term}.
Therefore, once the probe field reaches the vicinity of 
the event horizons, the contribution of the nonlinear dispersion 
terms cannot be neglected.
Combining the results of Eqs.~\eqref{eq:QNM result}, \eqref{eq:E_NL order},
Fig.~\ref{fig:QNM_WKBorder} and Table~\ref{table:QNM},
we expect that
it may be possible to observe
the behavior of a few QNM cycles 
experimentally,
since the QNM fundamental frequency is determined 
by the inverse of the timescale at which the probe field 
reaches the event horizons.

\section{Conclusion}
\label{sec:conclusion}

In this paper, we have investigated the perturbative 
stability of the circuit system of
TWPA with SNAILs.
We have derived the master equation for the weak probe field 
living on top of the background soliton solution.
We first have demonstrated 
that the master equation in the effective theory,
where the nonlinear dispersion is neglected, 
has no normalizable negative modes
corresponding to unstable solutions
by using the language of 
supersymmetric quantum mechanics.
We have also shown that 
the effective potential induced by the spatial modulation 
of the probe field due to the soliton can support the QNM and discussed the possibility of observing the ringdown
in the case of the SNAIL-TWPA 
analogue black-white holes.
We found that the QNM fundamental frequency 
can typically be evaluated as a function
of the normalized soliton relative velocity $\beta_{\rm phys}$
multiplied by the inverse of the timescale at which 
the probe field reaches the event horizons.
We have also considered the effects of the nonlinear dispersion term and examined when it becomes effective.
We found that for a few QNM cycles, the linear dispersion 
term dominates, suggesting the possibility of observing 
the QNM ringdown before nonlinear dispersion becomes important.
Once the probe field reaches the event horizons, 
the contribution of the nonlinear dispersion term can 
no longer be ignored.

Finally, we would like to discuss the subtleties involved 
in this paper.
In the numerical analysis to evaluate the QNM frequency, 
we have taken $|\beta_{\rm phys}|={\cal O}(0.1)$ so that
the correction terms in the WKB approximation are comparatively small.
However, when considering realistic circuit systems, 
the amplitude of solitons $A$ cannot be freely determined; 
it must be sufficiently small for the reductive perturbation method to remain valid. 
Furthermore, the value of the nonlinear parameters $c_{3,4}$ 
cannot naturally be chosen to be large.
In other words, in more realistic settings, we may need 
to choose a smaller value for $\beta_{\rm phys}$ 
than the one discussed in this paper.
Determining a more realistic value of $\beta_{\rm phys}$ 
is beyond the scope of this paper, 
but further investigation will be necessary in the future.

Let us also discuss the subtleties in 
the WKB approximation to the QNM of our
analogue black-white holes.
Since the potential itself also explicitly depends on $\epsilon$\,,
even if one takes the limit as $\beta_{\rm phys}=\epsilon\beta\to 0$ 
in the potential, 
$\beta_{\rm phys}$ is still contained within the definition 
of the $\eta^\ast$ 
derivatives. 
Consequently, the naive application of the (higher-order) 
WKB approximation may not yield the correct result.
In order to conduct a more precise evaluation, it is necessary 
to establish a methodology that can effectively apply 
more precise numerical calculation techniques, such as Leaver's method~\cite{Leaver:1985ax}.
However, the main purpose of this paper is to analytically evaluate 
the QNM frequency and discuss the validity 
of approximations based on the Klein-Gordon equation. 
Therefore, this is left for future research.

\section*{Acknowledgment}

We thank N. Hatakenaka, M. P. Blencowe, and S. Higashitani for their helpful discussions. This work is partly supported by JSPS KAKENHI Grant Numbers JP22K03627,
JP23K25868, JP25K21670 (D.Y.), JP21H05189, JP22H05111, JP25K07282 (N.T.), 25K17317 (H.K.), and by the HIRAKU Global Program (H.K.), funded by MEXT’s ``Strategic Professional Development Program for Young Researchers.''

\appendix

\section{Tortoise coordinate}
\label{sec:Tortoise coordinate}

In this Appendix, we show the explicit
form of the tortoise coordinate as
a function of $\eta$ and discuss
the QNM boundary conditions.
Substituting Eq.~\eqref{eq:v^2} into
Eq.~\eqref{eq:tortoise coordinate} and
integrating it, we have
\al{
    \eta^\ast =-\xi
    \Biggl[&
        \frac{1}{2}\sigma
        \log\left(\frac{\sqrt{\cosh^2(2\eta /w)+6\beta_{\rm phys}} +\sigma\sinh (2\eta /w)}{\sqrt{\cosh^2(2\eta /w)+6\beta_{\rm phys}} -\sigma\sinh (2\eta /w)}\right)
    \notag\\
    &
        -\log\left(\frac{\sqrt{\cosh^2(2\eta /w)+6\beta_{\rm phys}}-\sinh (2\eta/w)}{\sqrt{1+6\beta_{\rm phys}}}\right)
        \Biggr]
}
with
\al{
    \xi =\frac{w}{2v_0\beta_{\rm phys}(2+\beta_{\rm phys})}
    \,,\ \ \ 
    \sigma =\frac{\sqrt{6}(1+\beta_{\rm phys})}{\sqrt{4-\beta_{\rm phys}}}
    \,,
}
for the (a) KdV model, 
\al{
    \eta^\ast =-\xi
    \Biggl[&
        \frac{1}{2}\sigma\log
        \left(
            \frac{\sqrt{\cosh^2(2\eta /w)+12\beta_{\rm phys}}+\sigma\sinh (2\eta /w)}{\sqrt{\cosh^2(2\eta /w)+12\beta_{\rm phys}}-\sigma\sinh (2\eta /w)}
        \right)
    \notag\\
    &
    -\log\left(\frac{\sqrt{\cosh^2(2\eta /w)+12\beta_{\rm phys}}-\sinh (2\eta/w)}{\sqrt{1+12\beta_{\rm phys}}}\right)
        \Biggr]
}
with
\al{
    \xi =\frac{w}{2v_0\beta_{\rm phys}(2+\beta_{\rm phys})}
    \,,\ \ \ 
    \sigma =\frac{2\sqrt{3}(1+\beta_{\rm phys})}{\sqrt{10-\beta_{\rm phys}}}
    \,,
}
for the (b) mKdV$^+$ model, and
\al{
    \eta^\ast
        =-\xi
        \Biggl[&
        \frac{1}{2}\sigma\log
        \left(
            \frac{\sqrt{(1+6\beta_{\rm phys})\cosh^2(2\eta /w)-6\beta_{\rm phys}}+\sigma\sqrt{1+6\beta_{\rm phys}}\sinh (2\eta /w)}{\sqrt{(1+6\beta_{\rm phys})\cosh^2(2\eta /w)-6\beta_{\rm phys}}-\sigma\sqrt{1+6\beta_{\rm phys}}\sinh (2\eta /w)}
        \right)
    \notag\\
    &
        -\log\left(\sqrt{(1+6\beta_{\rm phys})\cosh^2(2\eta /w)-6\beta_{\rm phys}}+\sqrt{1+6\beta_{\rm phys}}\sinh (2\eta/w)\right)
        \Biggr]
    \,,
}
with
\al{
    \xi =\frac{w}{2v_0\beta_{\rm phys}(4-\beta_{\rm phys})}
    \,,\ \ \ 
    \sigma =\frac{\sqrt{6}(1+\beta_{\rm phys})}{\sqrt{2+\beta_{\rm phys}}}
}
for the (c) mKdV$^-$ model.
Here, we have chosen the boundary condition such that $\eta^\ast =0$ at $\eta =0$.
The position of the event horizons
in terms of $\eta$ is given by
$\eta = \pm \eta_{\rm H}$, where
\al{
    \eta_{\rm H}=\begin{dcases*}
    \frac{w}{2}{\rm arccosh}\left(\sqrt{\frac{6}{2+\beta_{\rm phys}}}\right)  & if (a)\,KdV\,, \\
    \frac{w}{2}{\rm arccosh}\left(\sqrt{\frac{12}{2+\beta_{\rm phys}}}\right) & if (b)\,mKdV$^+$\,, \\
    \frac{w}{2}{\rm arctanh}\left(\sqrt{\frac{2+\beta_{\rm phys}}{6}}\right)  & if (c)\,mKdV$^-$\,.
  \end{dcases*}
}
With these expressions, we find that, near the horizon, namely 
$\eta\to \pm \eta_{\rm H}$,
the tortoise coordinate can be well approximated as
\al{
    \eta^\ast
    \approx
    \mp
    \,\frac{1}{\kappa}\log |\eta \mp\eta_{\rm H}|
    +({\rm const.})
    \,,
}
where
\al{
    \kappa\equiv &\frac{2}{\xi\sigma}
    =\begin{dcases*}
    \frac{4v_0|\beta_{\rm phys}|(2+\beta_{\rm phys})\sqrt{4-\beta_{\rm phys}}}{\sqrt{6}w(1+\beta_{\rm phys})} & if (a)\,KdV\,, \\
    \frac{2v_0|\beta_{\rm phys}|(2+\beta_{\rm phys})\sqrt{10-\beta_{\rm phys}}}{\sqrt{3}w(1+\beta_{\rm phys})} & if (b)\,mKdV$^+$\,, \\
    \frac{4v_0|\beta_{\rm phys}|(4-\beta_{\rm phys})\sqrt{2+\beta_{\rm phys}}}{\sqrt{6}w(1+\beta_{\rm phys})}  & if (c)\,mKdV$^-$\,.
  \end{dcases*}
  \label{eq:kappa-def}
}
The QNM boundary conditions Eq.~\eqref{eq:bc} can be recast 
in terms of the $\eta$ coordinate as
\al{
    H\propto |\eta \mp \eta_{\rm H}|^{-{\rm i}\epsilon\Omega /\kappa}
    \ \ (\eta\to \pm \eta_{\rm H})
    \,.
}

\section{Details of the shooting method}
\label{sec:Details of the shooting method}

In this Appendix, we comment on some details of the shooting method used to compute the QNM frequencies.
We follow the procedure described in Sec.~\ref{sec:shooting method} to conduct the shooting method.
As in the main text, we take the convention $\eta_{\rm H} > 0$ for the position of the event horizons.

Instead of Eq.~\eqref{eq:Fourier KG}, we work on its dimensionless version given by
\al{
	\biggl\{\frac{\Delta\widetilde v^2}{\widetilde v^2}\frac{\dd}{\dd\widetilde \eta}\left[ \Delta\widetilde v^2\frac{\dd}{\dd\widetilde\eta}\right]+\widetilde \Omega^2\biggr\} \delta\varphi(\widetilde\eta) =0
	\,,\label{eq:dimensionless-KG}
}
where the dimensionless quantities in this equation are given by
\al{
\widetilde\eta &:= \eta / w\,, \\
\widetilde v^2 &:= v^2 / v_0^2\,, \\
\widetilde v_{\rm S} &:= v_{\rm S} / v_0 = 1 + \beta_{\rm phys} \,, \\
\Delta\widetilde v^2 &:= \bigl(\widetilde v^2 - \widetilde v_{\rm S}^2\bigr)/|\beta_{\rm phys}|\,, \\
\widetilde \Omega &:=
\frac{\epsilon w}{v_0|\beta_{\rm phys}|} \Omega\,.
}
The factors of $|\beta_{\rm phys}|$ in $\Delta\widetilde v^2$ and in
$\widetilde\Omega$ compensate each other, so that Eq.~\eqref{eq:dimensionless-KG}
is equivalent to Eq.~\eqref{eq:Fourier KG}.
The dimensionless frequency $\widetilde \Omega$ is related to $\Omega_{\rm phys}=\epsilon^{3/2}\Omega$ defined at Eq.~\eqref{eq:omega2_phys} as
\begin{equation}
\Omega_{\rm phys}=\epsilon^{3/2}\Omega
= \widetilde\Omega\times |\beta_{\rm phys}| \times \frac{v_0}{w_{\rm phys}}\,,
\label{eq:convert-to-Omega_phys}
\end{equation}
where $w_{\rm phys} = w / \sqrt{\epsilon}$ (defined at around Eq.~\eqref{eq:QNM result}).

\subsection{Details on the numerical method}

One subtlety in the construction of $\delta\varphi_\pm(\widetilde\eta)$ is that the wave equation~\eqref{eq:Fourier KG} has regular singular points at $\widetilde \eta =\pm \widetilde \eta_{\rm H} := \pm \eta_{\rm H} / w$\,.
This issue is resolved by using the Frobenius method to construct the series solutions near the horizons, that is,
\al{
    \delta\varphi_\pm (\widetilde \eta) 
    =
    |\widetilde \eta \mp \widetilde \eta_{\rm H}|^{-{\rm i}\widetilde\Omega /\widetilde\kappa}\sum_{n=0}^{n_{\rm max}} a_n^{(\pm)}|\widetilde\eta \mp \widetilde\eta_{\rm H}|^n
    \,,
}
where $a_n^{(\pm)}$ are the coefficients determined by solving the wave equation order by order in the series expansion for $\widetilde \eta\to \pm \widetilde \eta_{\rm H}$\,.
We also introduced the dimensionless version of $\kappa$ (Eq.~\eqref{eq:kappa-def}) as $\widetilde\kappa := \frac{w}{v_0 |\beta_{\rm phys}|} \kappa$\,.\footnote{%
$\widetilde \kappa$ is expressed also as
$\widetilde \kappa = \bigl(\dd \Delta\widetilde v^2 / \dd \widetilde \eta\bigr)/\widetilde v
= \bigl(\dd \widetilde v^2 / \dd \widetilde \eta\bigr)/\bigl(|\beta_{\rm phys}|\,\widetilde v\bigr)$
evaluated at the horizon $\widetilde\eta = - \widetilde \eta_{\rm H}$, as we can verify using Eqs.~\eqref{eq:tortoise coordinate} and \eqref{eq:H_BC}. We used this expression in the actual numerical code to automatically generate $\widetilde\kappa$ given from Eq.~\eqref{eq:kappa-def}.}

Using the above series solution, we specify the initial conditions $\delta\varphi_\pm, \dd \delta\varphi_\pm / \dd\widetilde\eta $ at $\widetilde\eta =\pm 
\widetilde \eta_{\rm H}\left(1 -\delta\right)$, where $\delta$ is a small positive number,
for the numerical integration of the wave equation toward $\widetilde\eta =0$\,.
As explained around Eq.~\eqref{eq:Wronskian-reduced} in Sec.~\ref{sec:shooting method}, we need to solve only for $\delta\varphi_\pm(\widetilde\eta)$ in $-\widetilde\eta_{\rm H}(1 - \delta) \leq \widetilde \eta \leq 0$ to determine the QNM frequency.
In our numerical calculations, we set $n_\text{max} = 16$ and $\delta=0.1$\,.

Equation~\eqref{eq:Wronskian-reduced} implies that the QNM frequency $\widetilde\Omega$ is determined by either $\delta\varphi_-(0)=0$ or $\dd\delta\varphi_- / \dd \widetilde\eta (0)=0$\,, each of which corresponds to the odd or even mode, respectively.
The left-hand sides of these equations are functions of $\widetilde\Omega$\,; their roots can be found by plotting the (complex) values of the left-hand sides as functions of $\widetilde\Omega$, and searching for the points where they become zero.
After identifying the approximate locations of the roots, we can use a root-finding algorithm, such as Mathematica's \texttt{FindRoot} command, to locate them more precisely.
The results summarized in Sec.~\ref{sec:5} are obtained from this procedure with the conversion to $\Omega_{\rm phys}$ according to Eq.~\eqref{eq:convert-to-Omega_phys}.

\subsection{Examples of numerical results}

As examples, we show some numerical results for the (a) KdV model with $\bp = 0.3$.
We particularly show the global structure of the roots of the Wronskian condition equation~\eqref{eq:Wronskian-reduced} on the complex plane, and also the roots corresponding to the lowest even and odd modes.
The results for the other models with different values of $\bp$, as well as for higher modes, are qualitatively similar to those.

\subsubsection{Global structure of the roots}

\begin{figure*}[tbp]
    \centering
    \subfloat[][$\dd \delta\varphi_- / \dd \widetilde\eta\,(0;\widetilde\Omega)$]{
    \includegraphics[height=7cm]{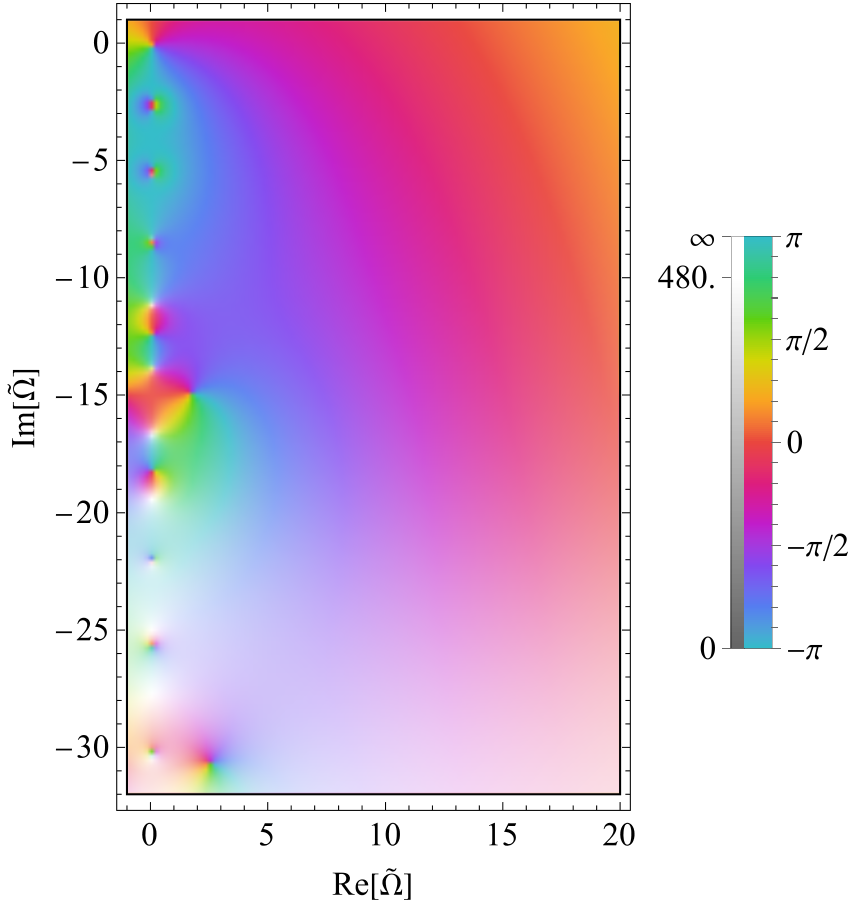}
    \label{fig:global_even}
    }
    \qquad
    \subfloat[][$\delta\varphi_-(0;\widetilde\Omega)$]{
    \includegraphics[height=7cm]{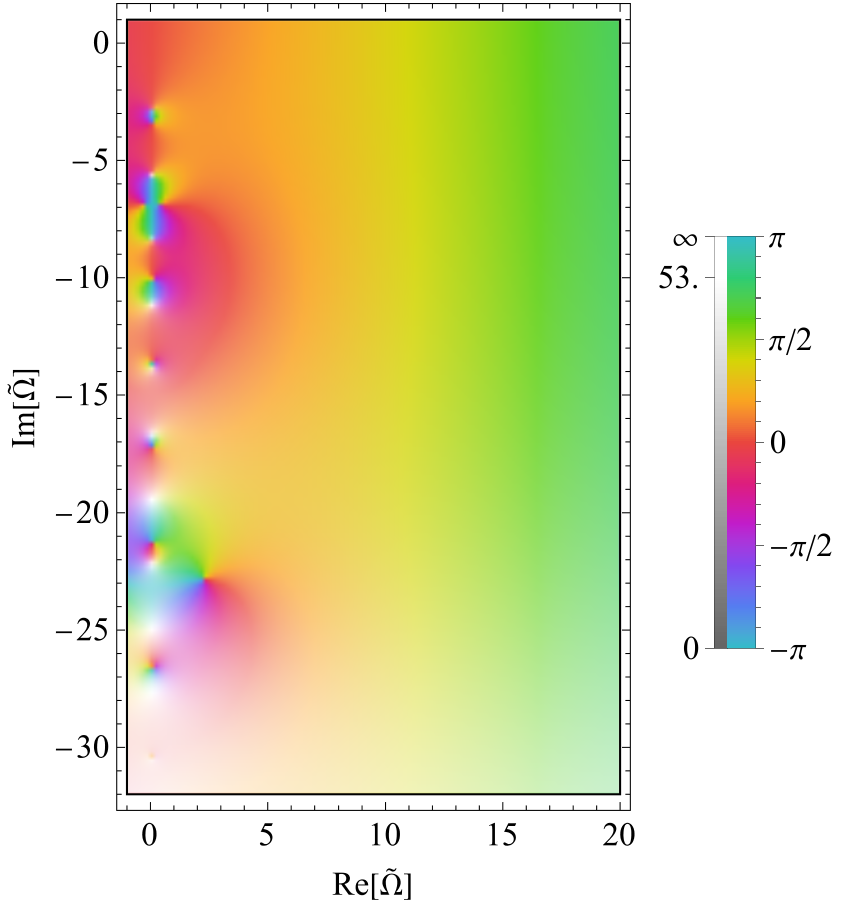}
    \label{fig:global_odd}
    }
   \caption[]{The complex values of $\dd\delta\varphi_-/\dd\widetilde\eta\,(0;\widetilde\Omega)$  (even modes, Fig.~\ref{fig:global}\subref{fig:global_even}) and $\delta\varphi_-(0;\widetilde\Omega)$ (odd modes, Fig.~\ref{fig:global}\subref{fig:global_odd}) on the complex $\widetilde\Omega$ plane for $\bp = 0.3$.
The hue and the color density represent the argument and absolute value, respectively. The roots (QNM frequencies) are located at the points with the highest color density. The region $\Re \widetilde\Omega < 0$ is omitted because the root positions are symmetric with respect to the imaginary axis.}
   \label{fig:global}
\end{figure*}

In Fig.~\ref{fig:global}, we show the complex values of $d\delta\varphi_-/d\widetilde\eta(0;\widetilde\Omega)$  (even modes, Fig.~\ref{fig:global}\subref{fig:global_even}) and $\delta\varphi_-(0;\widetilde\Omega)$ (odd modes, Fig.~\ref{fig:global}\subref{fig:global_odd}) on the complex $\widetilde\Omega$ plane for $\bp = 0.3$.
The hue and the color density represent the argument and absolute value, respectively. The roots (QNM frequencies) are located at the points of highest color density.

In Fig.~\ref{fig:global}\subref{fig:global_even}, we can find a root at $\widetilde \Omega = 0$. It corresponds to a trivial solution $\delta\varphi = $ constant, which does not describe a physical QNM mode. The other roots correspond to physical QNM frequencies.

In both Figs.~\ref{fig:global}\subref{fig:global_even} and \ref{fig:global}\subref{fig:global_odd}, we confirmed that the roots shown in the figures are robust against numerical errors, which strongly depend on the choice of the cutoff parameter $\delta$ and the order $n_\text{max}$ of the series solution used to set the initial conditions.
Neglecting the unphysical root at $\widetilde\Omega = 0$, the first few QNM frequencies are found on the negative part of the imaginary axis for both even and odd modes.

\subsubsection{Lowest modes}

\begin{figure}[tb]
    \centering
    \includegraphics[width=7.5cm]{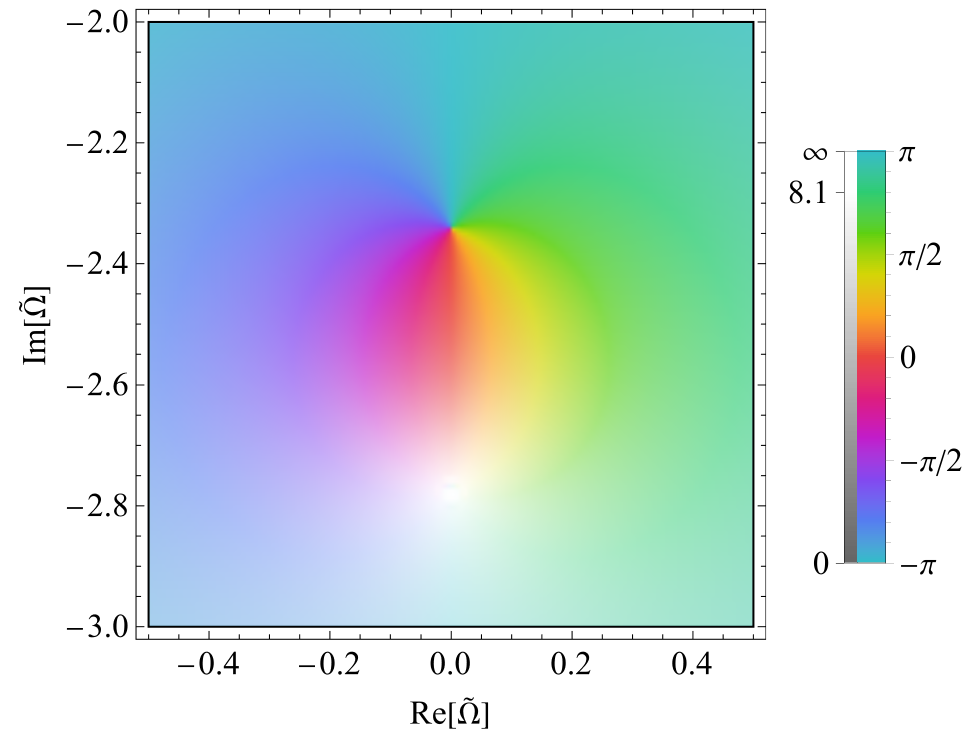}
    \caption{The close-up view of the root for the lowest even mode, at which $\dd\,\delta\varphi_-/\dd \widetilde\eta\,(0;\widetilde\Omega) = 0$ on the complex $\widetilde\Omega$ plane, for $\bp = 0.3$ in the (a) KdV model.}
   \label{fig:even_1st}
\end{figure}

We first analyze the lowest even mode, which corresponds to the root of $\dd\,\delta\varphi_-/\dd\widetilde\eta(0;\widetilde\Omega)$ with respect to $\widetilde\Omega$ with the largest imaginary part. 
In Fig.~\ref{fig:even_1st}, we show the close-up view of the root for the lowest even mode. From these numerical results, we can read off the QNM frequencies as $\widetilde\Omega = -2.34 \, {\rm i}$\,.
This frequency corresponds to $\Omega_{\rm phys} = -0.702 \,{\rm i} \times v_0 / w_{\rm phys}$ (see Eq.~\eqref{eq:convert-to-Omega_phys}) and is listed in Table~\ref{table:QNM}.

\begin{figure}[tb]
    \centering
    \includegraphics[width=7.5cm]{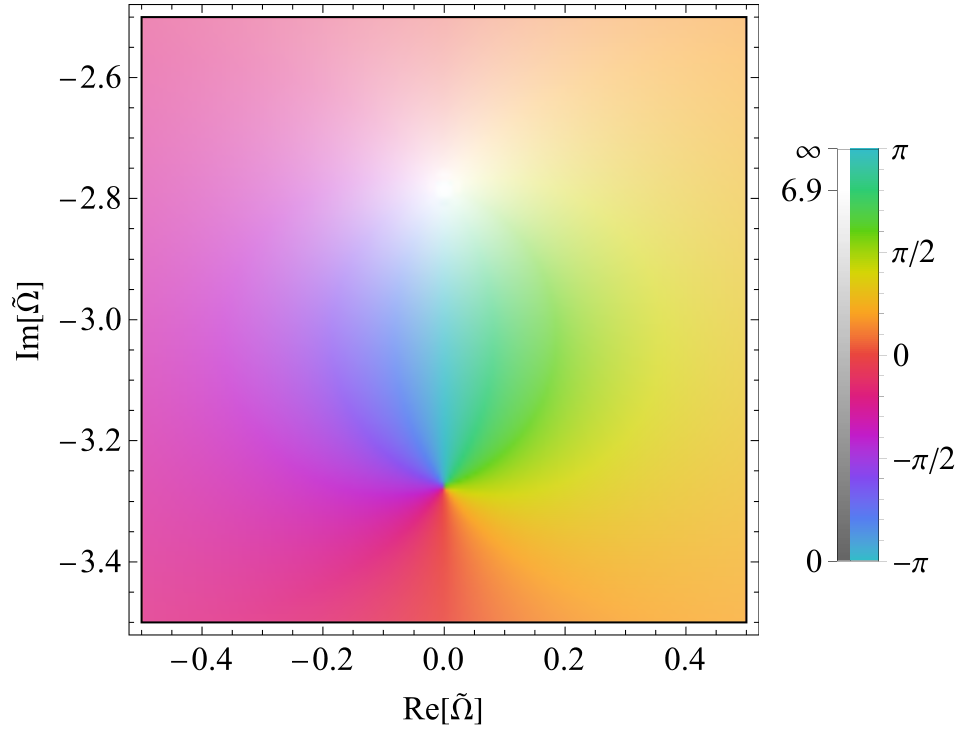}
   \caption{The close-up view of the root for the lowest odd mode, at which $\delta\varphi_-(0;\widetilde\Omega) = 0$ on the complex $\widetilde\Omega$ plane, for $\bp = 0.3$ in the (a) KdV model.}
   \label{fig:odd_1st}
\end{figure}

Next, we analyze the lowest odd mode, which corresponds to the root of $\delta\varphi_-(0;\widetilde\Omega) = 0$ with respect to $\widetilde \Omega$.
The close-up view of the roots for the lowest and the second odd modes is shown in Fig.~\ref{fig:odd_1st}. We can read off the QNM frequency as $\widetilde\Omega = -3.28 \, {\rm i}$ from this result.
This frequency corresponds to $\Omega_{\rm phys} = -0.983 \,{\rm i} \times v_0 / w_{\rm phys}$ (see Eq.~\eqref{eq:convert-to-Omega_phys}) and is listed in Table~\ref{table:QNM}.
Its imaginary part is smaller (more strongly damped) than that of the lowest even mode, as expected from the fact that the mode function $\delta\varphi(\widetilde\eta)$ of the lowest even and odd modes has zero and one node, respectively.

\subsection{Validation with the P\"oschl-Teller potential}

As a sanity check of the numerical method employed above, we apply the shooting method to the P\"oschl-Teller potential, for which the QNM frequencies are known analytically.
We observe that our numerical results accurately reproduce the exact solutions.

The Schr\"odinger equation with the P\"oschl-Teller potential is given by
\begin{equation}
    \left(
    \frac{\dd^2}{\dd x^2}
    + \omega^2
    - \frac{V_0}{\cosh^2 \alpha x}
    \right)
    \psi = 0\,,
    \label{PTeq}
\end{equation}
where $V_0$ and $\alpha$ are positive constants.
Imposing the outgoing boundary conditions at $x \to \pm \infty$, 
which is given by
\begin{equation}
    \psi \propto
    \begin{cases}
        e^{-{\rm i} \omega x} & (x \to -\infty) \\
        e^{+{\rm i} \omega x} & (x \to +\infty)        
    \end{cases}
    \label{PT-BC}
\end{equation}
the QNM frequencies are given by
\begin{equation}
    \omega = 
    \sqrt{V_0 - \frac{\alpha^2}{4}} - {\rm i}\alpha\left(n+\frac12\right)
    \qquad (n=0,1,2,\ldots)\,.
    \label{PT-omega}
\end{equation}

To apply our numerical method to this problem, we compactify the spatial coordinate from $x\in(-\infty,\infty)$ into $y\in (-1,1)$ by
\begin{equation}
   y=\tanh(\alpha x)\,,
   \qquad
   x = \frac{1}{2\alpha} \log\frac{1+y}{1-y}\,. 
\end{equation}
Then, 
the Schr\"odinger equation~\eqref{PTeq} and
the boundary conditions~\eqref{PT-BC} are rewritten as
\begin{equation}
    (1-y^2)^2 \frac{\dd^2\psi}{\dd y^2}
    - 2 y (1-y^2) \frac{\dd\psi}{\dd y}
    + \left(
    \frac{\omega^2}{\alpha^2}
    - \frac{V_0}{\alpha^2} (1-y^2)
    \right)
    \psi = 0\,,
    \label{PTeq_y}
\end{equation}
\begin{equation}
    \psi\propto
        \begin{cases}
        (1+y)^{-{\rm i}\omega / (2\alpha)} & (y \to -1) \\
        (1-y)^{-{\rm i} \omega / (2\alpha)} & (y \to +1)        
    \end{cases}
    \label{PT-BC_y}
\end{equation}

\begin{figure}[tb]
    \centering
     \includegraphics[width=7.5cm]{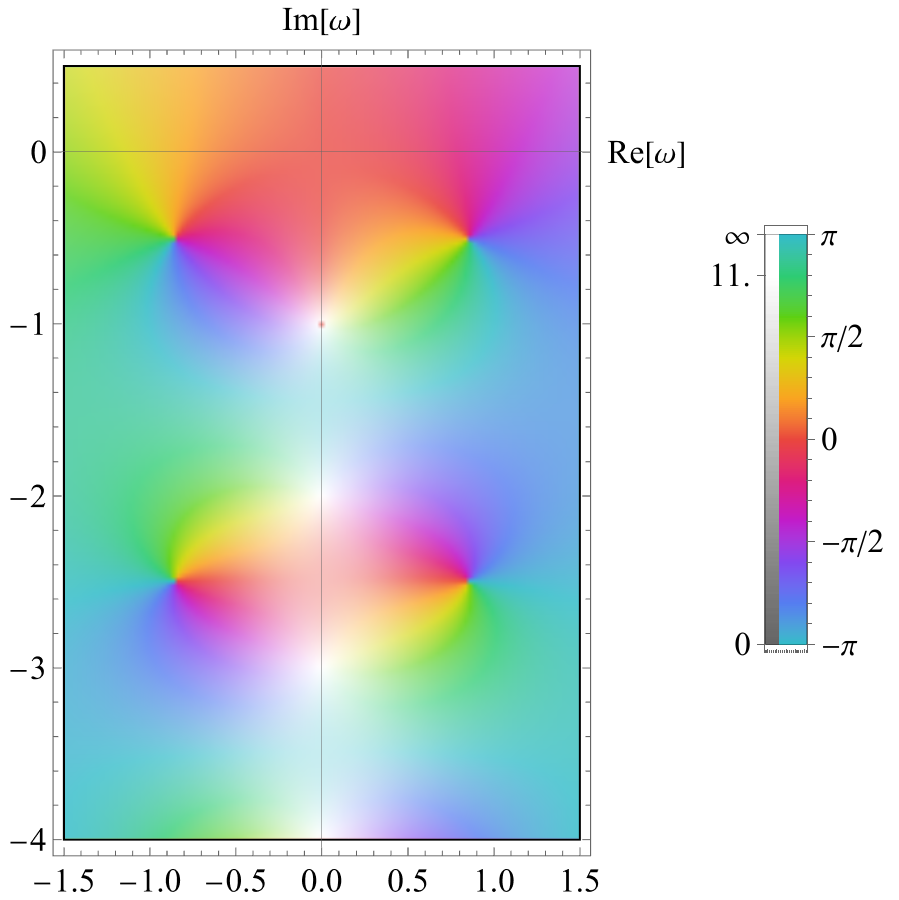}
    \caption{The complex value of $\dd\psi/\dd x\,(0; \omega)$ on the complex $\omega$ plane, whose zeros correspond to the QNM frequencies of the even modes, for the P\"oschl-Teller potential with $V_0 = \alpha = 1$.}
    \label{fig:PT_even}
\end{figure}

\begin{figure}[tb]
    \centering
     \includegraphics[width=7.5cm]{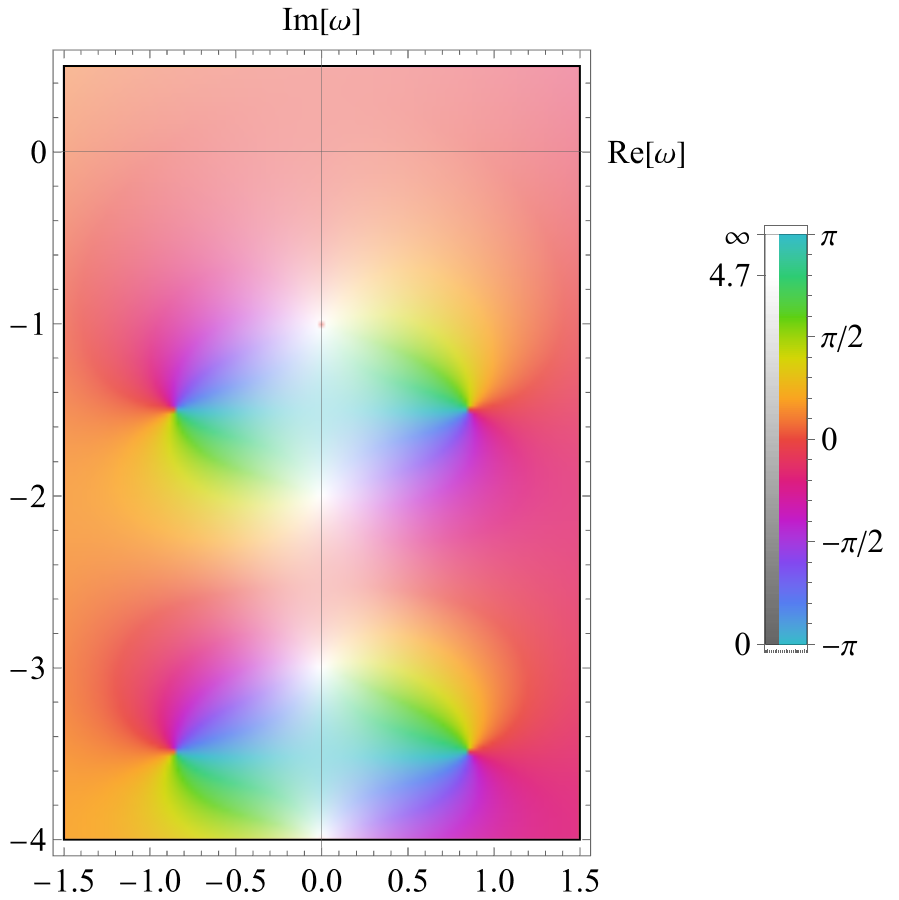}
    \caption{The complex value of $\psi(0; \omega)$ on the complex $\omega$ plane corresponding to the odd modes for the P\"oschl-Teller potential with $V_0 = \alpha = 1$.}
    \label{fig:PT_odd}
\end{figure}

In Figs.~\ref{fig:PT_even} and \ref{fig:PT_odd}, we show the distribution of the roots of the Wronskian on the complex $\omega$ plane for the P\"oschl-Teller potential with $V_0 = \alpha = 1$.
We can confirm that the roots obtained numerically agree well with the exact eigenvalues~\eqref{PT-omega} for $V_0 = \alpha = 1$, which are given by
\begin{equation}
    \omega = \pm 0.866\ldots - \left(n+\frac12\right){\rm i}\,,
    \qquad
    n=0,1,2,\ldots \, .
\end{equation}
For example, for the lowest (even) mode, the relative error of the numerical value of the frequency from the exact one (Eq.~\eqref{PT-omega}) for the lowest mode was $\mathcal{O}(10^{-7})$ for both real and imaginary parts when we use the same numerical setting as that used for the main problems.

In the examples in the main text, we observed that many pure imaginary modes appear. To examine if our numerical method properly works even for such pure imaginary modes, we conducted a validation for the case with $V_0 = 1/8, \alpha = 1$ that develops pure imaginary modes. We confirmed that the exact frequencies are
reproduced with a relative error $\mathcal{O}(10^{-8})$ at least for the first two lowest modes. These checks confirm that the shooting results are reliable for both complex and pure imaginary QNM frequencies.

\begin{figure*}[tb]
    \centering
    \includegraphics[height=45mm]{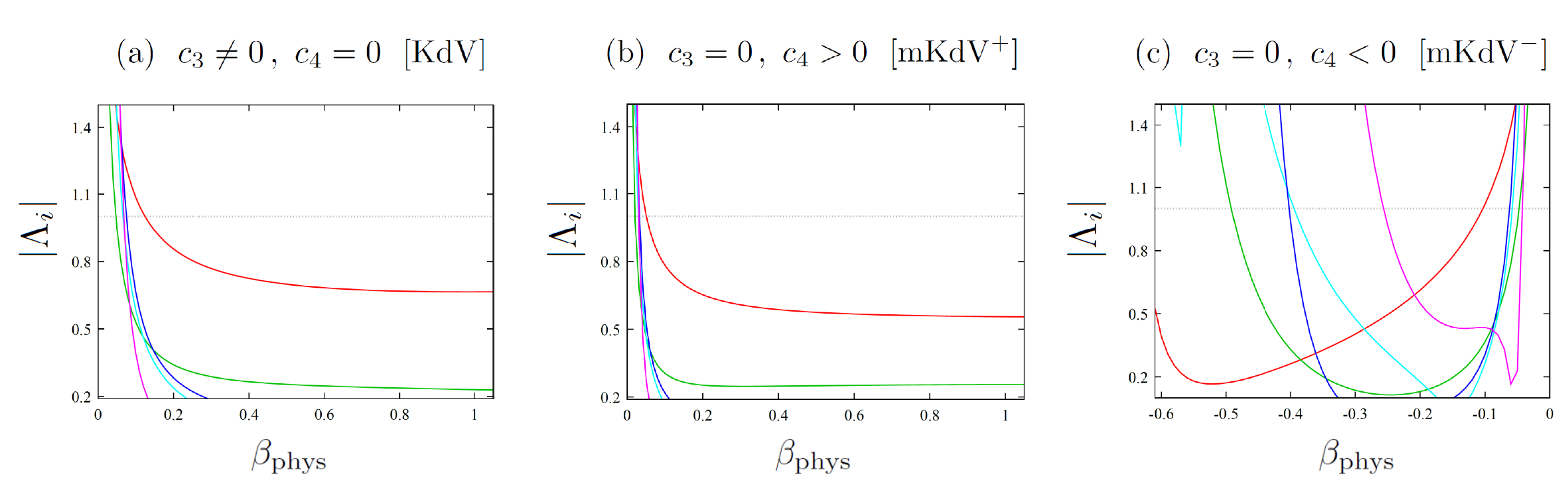}
     \caption{
     Second [red], third [green], fourth [cyan],
     fifth [blue], and sixth [magenta]
     order correction terms of the WKB approximation for KdV [left panel], mKdV$^+$ [center panel], and mKdV$^-$ [right panel] models.}
     \label{fig:WKB_validity}
\end{figure*}

\section{Validity of WKB approximation}
\label{sec:Validity of WKB approximation}

In this Appendix, we estimate the valid region
of the parameters for the WKB approximation 
to evaluate the QNM frequency in our system.
The WKB formula for the QNM has the form
\al{
	&\frac{{\rm i}(\epsilon^2\Omega^2-\widetilde V_0)}{\sqrt{-2\widetilde V_0^{\prime\prime}}}
    -\sum_{i=2}\Lambda_i =n+\frac{1}{2}
	\,,\label{eq:WKB-Lambda}
}
where the correction terms $\Lambda_i$ depend on the
value of the effective potential and its derivatives 
at the maximum.
The explicit form of the WKB corrections can be found 
in Refs~\cite{Iyer:1986np,Konoplya:2019hlu}
for $\Lambda_{2,3}$ and in Ref.~\cite{Konoplya:2003ii}
for $\Lambda_{4,5,6}$.

We use these expressions to evaluate the valid range of the WKB approximation.
As the most optimistic estimate, we define a valid range as one in which 
the absolute value of the higher-order terms, up to the sixth order, does not 
exceed that of the lowest-order term. 
In other words,
the correction terms should be suppressed compared
to the leading terms, in particular $|\Lambda_i| <n+1/2$, that is,
$|\Lambda_i|<1/2$ for the fundamental mode.
Therefore, we examine the parameter region
where the correction terms $\Lambda_i$
become smaller than $1/2$
to verify the validity of the WKB approximation.
We show the second [red], third [green], fourth [cyan], fifth [blue],
and sixth [magenta] order correction terms of
the WKB approximation for each model in Fig.~\ref{fig:WKB_validity},
where the dotted line indicates $|\Lambda_i|=1$.
This figure implies that 
in all cases, when $|\beta_{\rm phys}|$ is sufficiently small, 
such as $0.1$ or less, $\Lambda_i$ becomes 
greater than unity, 
indicating that the validity of the WKB approximation 
gets worse.
In comparison,
at least for $\beta_{\rm phys}\gtrsim 0.3$ for 
(a) KdV and (b) mKdV$^+$, and for $-0.2\lesssim\beta_{\rm phys}\lesssim -0.15$ 
for (c) mKdV$^-$\,
all $|\Lambda_i|$ are less than unity, and the correction terms are
comparatively small.
Even in these ranges, however, $|\Lambda_2|$ exceeds $1/2$, and at every value
of $\beta_{\rm phys}$ at least one of the $|\Lambda_i|$ does so.
The WKB approximation is therefore not well justified for any $\beta_{\rm phys}$
in our system, and we adopt the above ranges in the main text as those in which
its correction terms are comparatively small.

\vfill


\end{document}